\documentclass[fp,twocolumn]{jpsj3}
\usepackage{graphicx}
\usepackage{dcolumn}
\usepackage{bm}
\usepackage{subfigure}
\usepackage{xfrac}
\usepackage[utf8x]{inputenc}
\usepackage{color}
\usepackage{textcomp}
\newcommand\newblock{\ }

\newcommand{\op}[1]{%
    \fontdimen12\textfont3=2pt\fontdimen12\scriptfont3=1.4pt%
    \!\null\mathop{\vphantom{#1}\smash{#1}}\limits_{\sim}\null\!}
\newcommand{\xref}[1]{\protect\ref{#1}}
\newcommand{\figref}[1]{Fig.~\protect\ref{#1}}

\newcommand{\fmref}[1]{(\protect\ref{#1})}
\newcommand\mydots{\hbox to 1em{.\hss.\hss.}}

\def\ket#1{\, | \, {#1} \, \rangle}
\newcommand{\braket}[2]{\langle \, {#1} \, | \, {#2} \, \rangle}

\newcommand{\kagome}{kagom\'e\ }

\renewcommand{\eqref}[1]{Eq.~(\protect\ref{#1})}

\newcommand*{\subt}[1]{_{\text{\scriptsize #1}}}

\newcommand{\Hi}[1]{\mathcal{H}_{#1}}
\renewcommand{\i}{\textbf{i}}
\newcommand{\otp}[1]{($\sfrac{1}{3}{#1}$)-step}

\title{Melting of Magnetization Plateaus for Kagom\'e\ and 
Square-Kagom\'e\ Lattice Antiferromagnets}

\author{Henrik Schl\"uter$^1$\thanks{ORCID 0000-0001-5367-9879}, Johannes Richter$^2$\thanks{ORCID 0000-0002-5630-3786}, and J\"urgen Schnack$^1$\thanks{ORCID 0000-0003-0702-2723}}
\inst{$^1$Fakult\"at f\"ur Physik, Universit\"at Bielefeld, Postfach 100131, D-33501 Bielefeld, Germany \\
$^2$Institut f\"ur Theoretische Physik, Universit\"at Magdeburg, P.O. Box 4120, D-39016 Magdeburg, Germany \& Max-Planck-Institut f\"{u}r Physik Komplexer Systeme,
        N\"{o}thnitzer Stra{\ss}e 38, 01187 Dresden, Germany} 

\abst{Unconventional features of the magnetization curve at zero temperature such as plateaus or jumps
are a hallmark of frustrated spin systems. Very little is known about their behavior at non-zero temperatures. 
Here we investigate the temperature dependence of the magnetization curve of the \kagome lattice antiferromagnet 
in particular at $\sfrac{1}{3}$ of the saturation magnetization for large lattice sizes of up to $N=48$ spins. 
We discuss the phenomenon of asymmetric melting 
and trace it back to a combined effect of unbalanced magnetization steps on either side of the 
investigated plateau as well as on the behavior of the density of states across the plateau.
We compare our findings to the square-kagome lattice that behaves similarly at low temperatures 
at zero field, but as we will demonstrate differently at $\sfrac{1}{3}$ of the saturation magnetization.
Both systems possess a flat one-magnon band and therefore share with the class of flat-band systems 
the general property that the plateau that precedes the jump to saturation melts asymmetrically but now
with a minimal susceptibility that bends towards lower fields with increasing temperature.}

\begin{document}
\maketitle

\section{Introduction}
\label{sec-1}

Among the frustrated spin lattices the spin-$\sfrac{1}{2}$ \kagome\ Heisenberg antiferromagnet (KHAF) 
is one of the most prominent and at the same time ``enigmatic" spin
systems \cite{LSM:PRB19}. Practically all aspects of its magnetic properties are under debate:
(a) the precise nature of the spin-liquid ground state 
\cite{Yan2011,IBP:PRB11,DMS:PRL12,Laeuchli2011,Iqbal2013,Nor:RMP16,Pollmann2017,Xie2017},
(b) the magnetic and caloric properties at non-zero temperature
\cite{elstner1994,NaM:PRB95,ToR:PRB96,WEB:EPJB98,Lhuillier_thermo_PRL2000,Yu2000, 
Bernhard2002,Bernu2005,MiS:EPJB07,RBS:PRE07,Lohmann2014,Shimokawa2016,ShS:PRB18,
MZR:PRB18,CRL:SB18},
and (c) the magnetization process of the spin-$\sfrac{1}{2}$
KHAF \cite{Hida2001,SHS:PRL02,HSR:JP04,derzhko2004finite,zhitomirsky2004exact,
CGH:PRB05,SaN:PRB11,GMM:JPCM11,Nishimoto2013,CDH:PRB13,NaS:JPSJ14,KPO:PRB16,PMH:PRB18,NaS:JPSJ18,HCG:PB05,
DRH:LTP07,GeS:PRB22,Yos:JPSJ22}.

In the present paper we discuss the temperature dependence of the magnetization curve
of the KHAF for large system sizes.
At $T=0$ and in the thermodynamic limit the magnetization curve  
consists of a series of magnetization 
plateaus at $\sfrac{3}{9}=\sfrac{1}{3}$, $\sfrac{5}{9}$ and $\sfrac{7}{9}$ (and possibly \cite{Nishimoto2013} at $\sfrac{1}{9}$) of the
saturation magnetization \cite{Nishimoto2013,CDH:PRB13,CRL:SB18}, among which
the plateau at $\sfrac{1}{3}$ is the widest \cite{Hida2001}.
In the following we distinguish between plateaus that survive in the thermodynamic limit and magnetization 
steps that naturally arise due to the finite size of the investigated system.
It was noted in Ref.~\cite{SSR:PRB18} that the $\sfrac{1}{3}$-plateau which is flat at $T=0$ ``melts" 
rather quickly with increasing temperature and does so in an asymmetric way due to 
an unevenly balanced density of states across the plateau. 
Recent investigations on systems of size $N=27$ and $N=36$ 
confirm these findings and argue alongside \cite{MMY:PRB20}.
Sakai and Nakano even speculate about a magnetization ramp in the thermodynamic limit \cite{SaN:JKPS13}.

Here we investigate the matter in depth for large systems sizes of up to $N=48$ sites.
These results are obtained by large-scale numerical calculations using the
finite-temperature Lanczos method
(FTLM) \cite{JaP:PRB94,HaD:PRE00,ADE:PRB03,ScW:EPJB10,SuS:PRL12,SuS:PRL13,ScT:PR17,PrK:PRB18,SSR:PRB18,OAD:PRE18,IMN:IEEE19,MoT:PRR20}.
We discuss the behavior of the density of states in the vicinity of the 
$\sfrac{1}{3}$-plateau as well as the finite-size scaling of 
the neighboring 
magnetization steps. These steps influence the asymmetry as well.

Finally, we compare our findings with the magnetization curve of the related
square-kagome lattice Heisenberg antiferromagnet (SKHAF) \cite{RST:CMP09,NaS:JPSJ13,RDS:PRB22}, 
that does not exhibit 
asymmetric melting of the $\sfrac{1}{3}$-plateau.

The paper is organized as follow. In Section \ref{sec-2} we
introduce the model and our numerical scheme. Thereafter in
Section~\ref{sec-3} we present our results for the KHAF and the SKHAF followed
by a discussion in Section~\ref{sec-4}.

\section{Method}
\label{sec-2}
In this paper we use FTLM data to determine thermodynamic observables 
such as the magnetization $M(T, h)$ and the differential susceptibility $\chi(T, h)$ 
as well as 
the density of states $\rho(E, h)$. We employ the open-source software
\verb#spinpack#
of J{\"o}rg Schulenburg \cite{spin:258}.
The spin systems at hand are defined by the Hamiltonian
\begin{align}
\label{E-2-1}
    \op H = J 
    \underset{(i,j) \in \text{bonds}}{\sum\op{\vec{s}}_i\cdot\op{\vec{s}}_j}
    \ ,
\end{align}
where the set ``bonds" contains all pairs of connected sites $(i,j)$ of a lattice, e.g., 
nearest neighbors for the investigated kagome and square-kagome lattices. 
$J$ is called coupling constant and describes an antiferromagnetic coupling for $J>0$.
Due to the rotational -- SU(2) -- symmetry of the Heisenberg model, \eqref{E-2-1}, 
the orthogonal subspaces associated with total magnetic quantum number $M$ 
can be treated separately 
\begin{align}
    \Hi{}=\bigoplus_{M=M_{\text{min}}}^{M_{\text{max}}} \Hi{M}\ ,
\end{align}
where the sum runs over the orthogonal subspaces $\Hi{M}$.
The Zeeman term, added to \fmref{E-2-1}, contains a dimensionless magnetic field $h$ that relates 
to the magnetic flux density $B$ via $h = g \mu_B B$.

When constructing the density of states
\begin{align}
\rho(E, h)
=
\sum_{M=M_{\text{min}}}^{M_{\text{max}}} \rho_M(E, h)
\end{align}
from FTLM data, it should be noted that there is some freedom in smoothening it.
This problem exists already for the exact density of states, but is worse
for FTLM data that consists of much fewer discrete Lanczos energy eigenvalues.
In this work, we choose a representation where the pseudo-gaps in parts of the spectrum
that should be dense are avoided. A detailed description of the calculation
is given in the Appendix. 

The partition function approximated with the finite-temperature Lanczos method is given by 
\begin{align}
    Z(T, h)  = \sum_{M=M_{\text{min}}}^{M_{\text{max}}}
    \sum^{N_L}_{n=1}
    \sum^{R}_{r=1}
    \gamma_n^{(r, M)}\;
    e^{-\beta (\epsilon^{(r,M)}_n + h M)}
    \ ,
\end{align}
where $N_L$ is the number of steps in the Krylov space expansion and $R$ the number of random vectors in the 
typicality approach to approximate traces \cite{SRS:PRR20}. $\beta=1/(k_B T)$ denotes the inverse temperature. 
The Lanczos weights are
\begin{align}
    \gamma_n^{(r, M)} = 
        \frac{\text{dim}\Hi{M}}{R}
        \vert \braket{n(r,M)}{r,M}\vert^2
        \ ,
\end{align}
where $\ket{n(r,M)}$ is the normalized $n$-th eigenstate of the Krylov space expansion of the 
Hamiltonian with the initial state $\ket{r,M}$ and $\epsilon^{(r,M)}_n$ is the Krylov-space energy eigenvalue. 

All observables of interest can be derived from the partition function $Z(T, h)$. 
For large systems with $N>42$ the partition function is incomplete since 
some subspaces $\Hi{M}$ for small $|M|$ are too large for a Lanczos 
procedure. Such partition functions can still be used as accurate 
approximations at high enough fields and low temperatures. 
For the KHAF we take the following subspaces into account: 
$N=45: |M|\geq 3.5$,  
$N=48: |M|\geq 6$,  
$N=54: |M|\geq 18$,  
$N=63: |M|\geq 22.5$,
$N=72: |M|\geq 26$; 
and for the SKHAF: 
$N=48: |M|\geq 11$,  
$N=54: |M|\geq 15$,  
$N=60: |M|\geq 18$.
In the following we use the reduced temperature $t=k_B T/|J|$.


\section{Numerical results}
\label{sec-3}

In this section we investigate the influence of subspaces $\Hi{M}$ 
belonging to neighboring magnetization steps on the asymmetric melting of 
the $\sfrac{1}{3}$-plateau. We find that both the width of these steps
as well as the density of states of the related subspaces $\Hi{M}$ play
a role.

From Ref.~\citen{RDS:PRB22} it is known that the thermodynamic properties of the KHAF and 
the SKHAF at zero magnetic field are very similar.
Here we will demonstrate that the melting of the $\sfrac{1}{3}$-plateau
is significantly different in both systems.

\subsection{Asymmetric melting of the KAHF $\sfrac{1}{3}$-plateau}
\label{sec-3-1}

\begin{figure}[ht!]
    \centering
    \includegraphics[width=0.90\columnwidth]{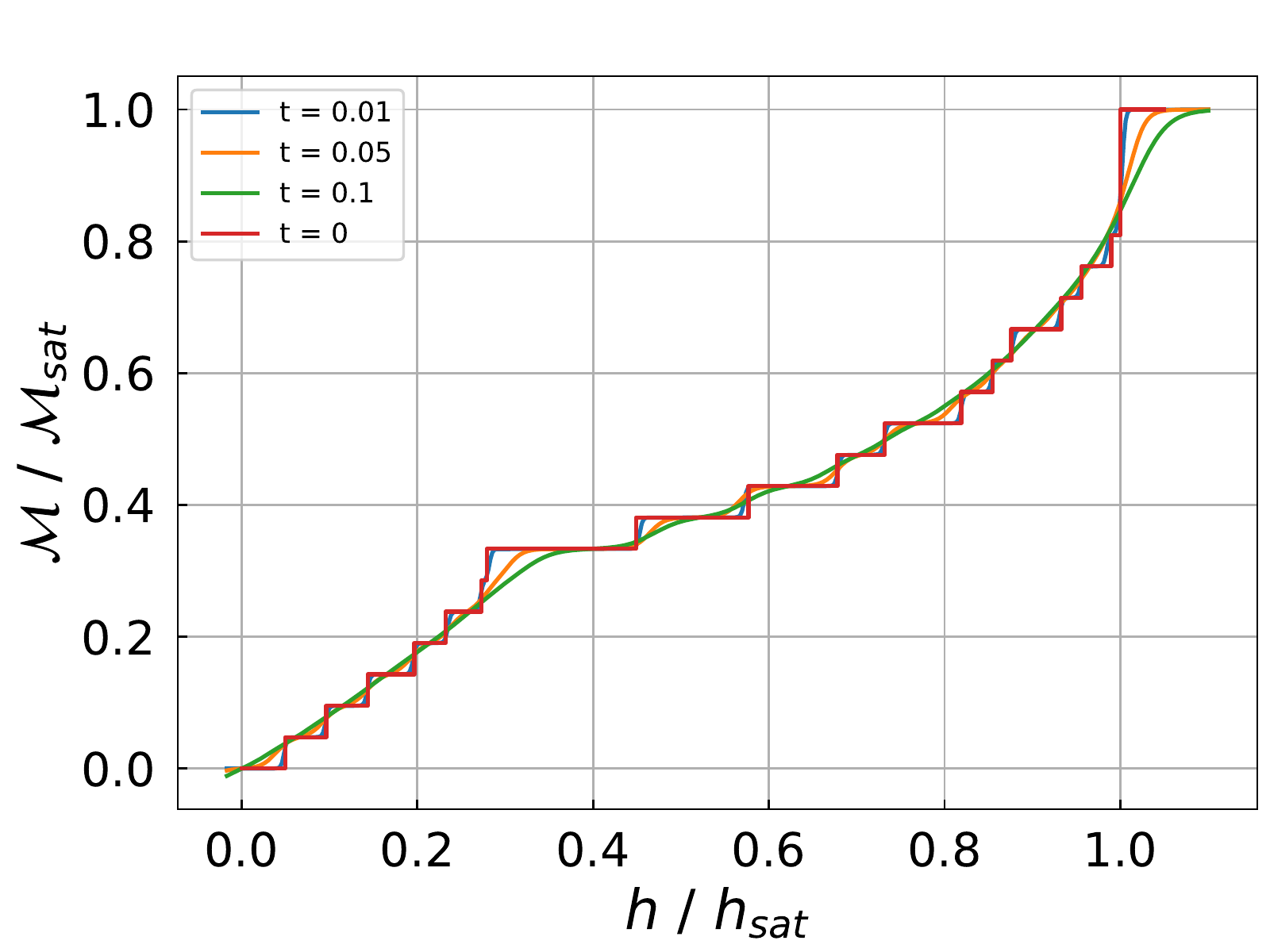}
    
    \includegraphics[width=0.90\columnwidth]{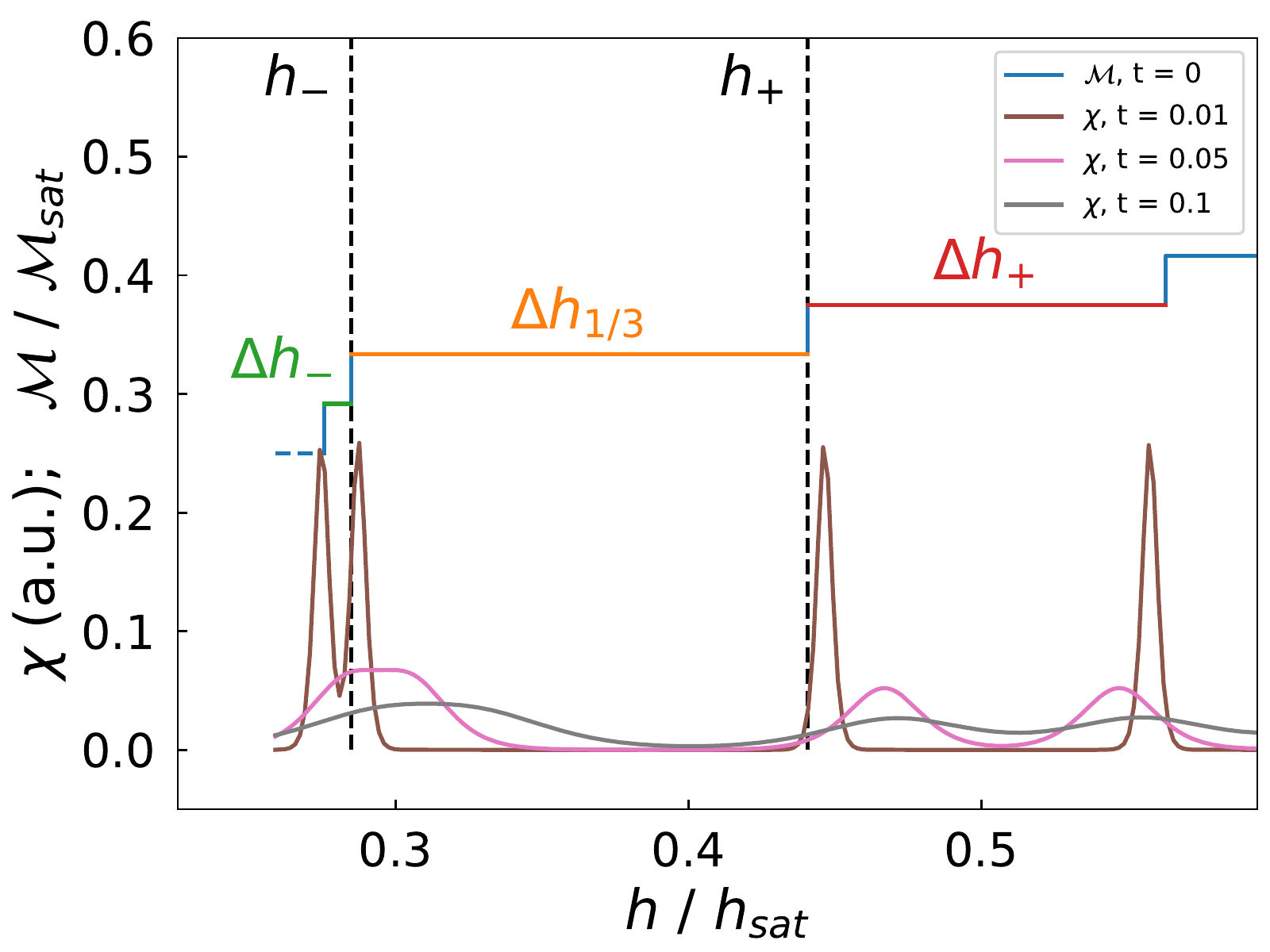}
    \caption{(Color online) Magnetization curves of the \kagome lattice with $N=42$ sites (top) 
    as well as with $N=48$ sites (bottom) for various temperatures together
    with the differential susceptibility for $N=48$ (bottom). 
    The asymmetric melting of the plateau at $\sfrac{1}{3}$ 
    of the saturation magnetization is clearly visible, 
    compare also \cite{SSR:PRB18}. $h_{-}$ and $h_{+}$ define the end
    points of the $\sfrac{1}{3}$-plateau; $\Delta h_{1/3}$ 
    denotes its width, 
    whereas $\Delta h_-$ and $\Delta h_+$ are the widths of the neighboring 
    magnetization steps.}
    \label{f-3-1}
 \end{figure} 

\paragraph{Magnetization curve:} 
The afore-mentioned asymmetric melting of the $\sfrac{1}{3}$-plateau
can be quantified by comparing the thermal behavior at the low-field 
end $h_{-}$ and at the high-field end $h_{+}$ of the plateau, see \figref{f-3-1}.
One can see that for increasing temperatures the magnetization at $h_-$ drops
rapidly from the zero-temperature value of $\sfrac{1}{3}$ whereas
the value of magnetization at $h_+$ roughly stays the same even for higher temperatures. 
A suggested explanation of this asymmetric phenomenon is that 
the density of states at low energies of the $M_{1/3}$-subspace is 
far denser than that of the $M_{1/3}+1$-subspace and that the density of states 
of the $M_{1/3}-1$-subspace must be even denser \cite{SSR:PRB18,MMY:PRB20}. 

As discussed in recent articles \cite{SSR:PRB18,MMY:PRB20} and shown later on, 
this explanation is partially correct,
but there is an additional cause for this phenomenon to be mentioned. 
As can be seen in \figref{f-3-1}, the \otp{-1} is very small, 
especially compared to the \otp{+1}. This suggests that for low temperatures where 
at $h_+$ only states from two subspaces contribute significantly to the magnetization, 
at $h_-$, states from three or four subspaces are involved. 
In Ref.~\citen{MMY:PRB20}, the possible influence of additional 
subspaces is acknowledged but not further investigated. 

\begin{figure}[ht!]
    \centering
    \includegraphics[width=0.9\columnwidth]{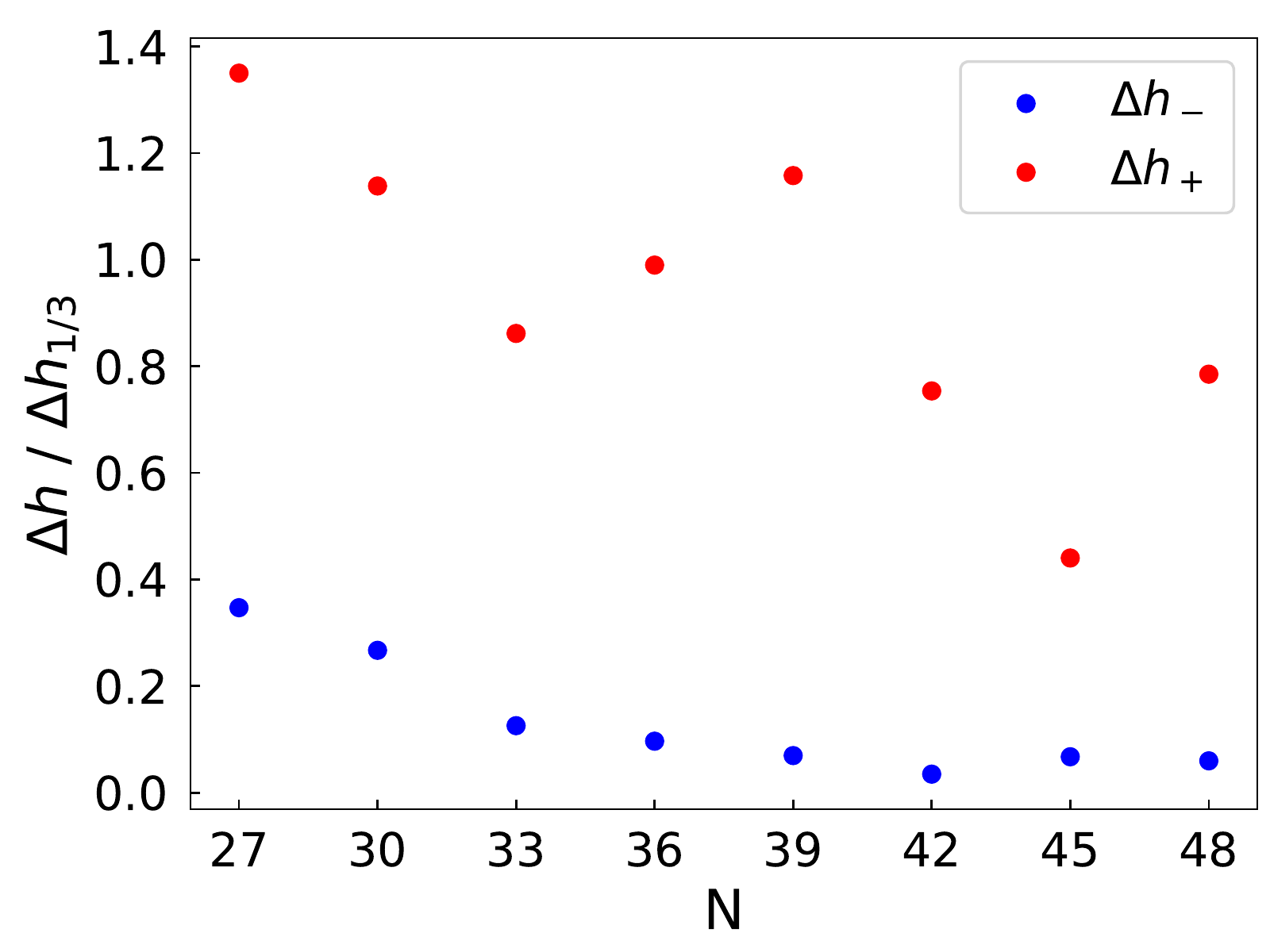}
    
    \includegraphics[width=0.9\columnwidth]{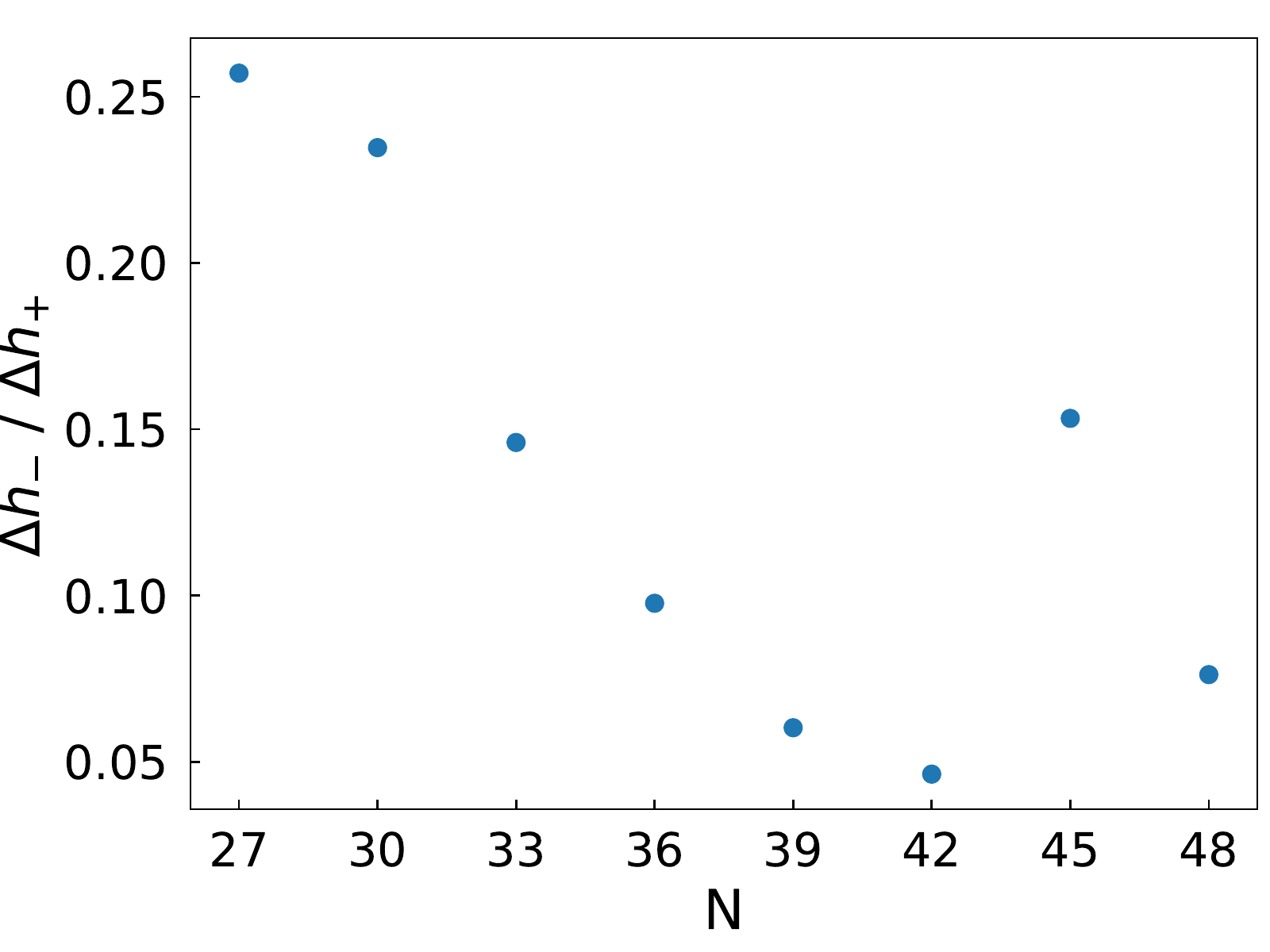}
    \caption{(Color online) The widths of the \otp{-1} $\Delta h_-$ and of the \otp{+1} $\Delta h_+$, 
    respectively. Top: normalized values, bottom: ratio $\Delta h_-/\Delta h_+$.}
    \label{f-3-2}
\end{figure}

Figure \xref{f-3-2} demonstrates that the step size $\Delta h_+$ 
is (typically much) greater than $\Delta h_-$
for all investigated system sizes. Even though the details vary due to finite size effects
it is evident that the width $\Delta h_-$ of the low-field step is typically
significantly less than a fifth of the width $\Delta h_+$ of the high-field magnetization step,
compare r.h.s. of \figref{f-3-2}.
The differences between $N=45$ and $N=42, 48$ in \figref{f-3-2} could be related to
the nature of the $\sfrac{1}{3}$-plateau state that can be understood as a 
valence-bond state with a magnetic unit cell of 9 spins 
\cite{CGH:PRB05,Nishimoto2013,CDH:PRB13,CRL:SB18}
and therefore fits much better to $N=45$ than to $N=42$ or $48$.

In order to investigate which subspaces $\Hi{M}$ contribute dominantly to the
magnetization at low temperatures, we compare the deviations from the exact 
value when estimating the magnetization with only subsets of the subspaces
$\Hi{M}$. 
To this end we define the following subsets:
\begin{align}
    \Gamma_2& \left(h_-\right) := \{M_{1/3}, M_{1/3} - 1\}\\
    \Gamma_3&\left(h_-\right) := \{ M_{1/3} ,  M_{1/3}  - 1,  M_{1/3} - 2\}
\end{align}
to be used for $\mathcal{M}(T, h_-)$. Here $\Gamma_2\left(h_-\right)\subset\Gamma_3\left(h_-\right)$, 
i.e., $\Gamma_2\left(h_-\right)$ yields
a more restrictive approximation of the magnetization.

To determine similar deviations at $h_+$ we consider the subsets 
\begin{align}
    \Gamma_2&\left(h_+\right) := \{M_{1/3} + 1, M_{1/3}\}\\
    \Gamma_3&\left(h_+\right) := \{ M_{1/3} + 2,  M_{1/3} + 1,  M_{1/3}\}
    \ ,
\end{align}
where $\Gamma_2\left(h_+\right)\subset\Gamma_3\left(h_+\right)$.
The subspace associated with these sets is defined as
\begin{align}
    \Hi{}(\Gamma_k(h)) := \bigoplus_{M\in\Gamma_k(h)}\Hi{M}\ .
\end{align}
The sets $\Gamma_k(h)$ are chosen such that $\Hi{}(\Gamma_k(h))$ contains 
the $k$ lowest-lying subspaces for the applied magnetic field $h$.   
We define the deviation
\begin{align}
    \label{e-3-1}
    \Delta \mathcal{M}_{\Gamma_i}(T, h) = \big\vert \mathcal{M}(T, h) - \mathcal{M}_{\Gamma_i}(T, h)\big\vert
    \ ,
\end{align}
where $\mathcal{M}_{\Gamma_i}(T, h)$ denotes the approximation of the 
magnetization using only the subspace $\Hi{}(\Gamma_i(h))$\ .

\begin{figure}[ht!]
    \vspace{0.3cm}
    \centering
    \includegraphics[width=0.9\columnwidth]{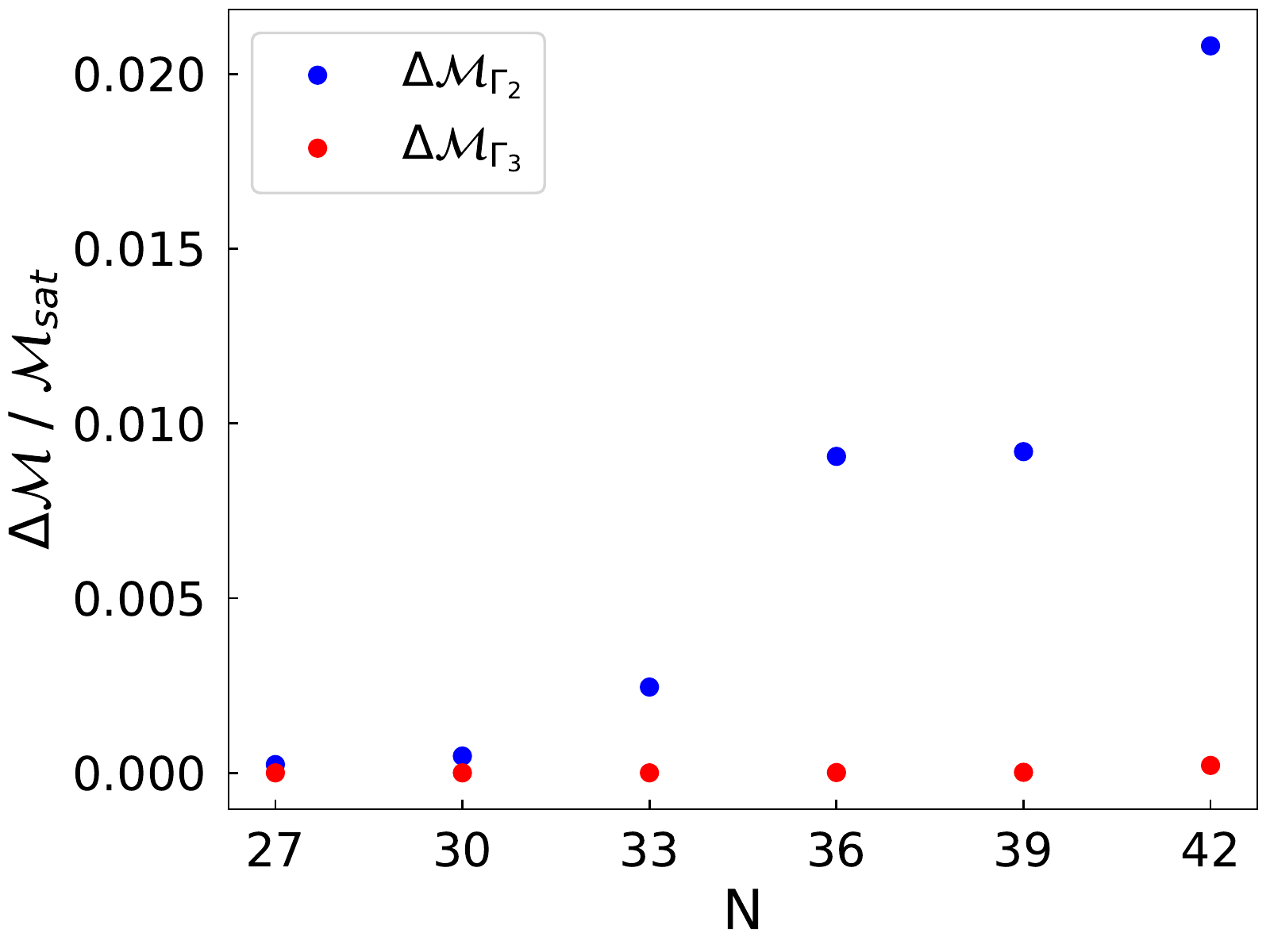}
    
    \includegraphics[width=0.9\columnwidth]{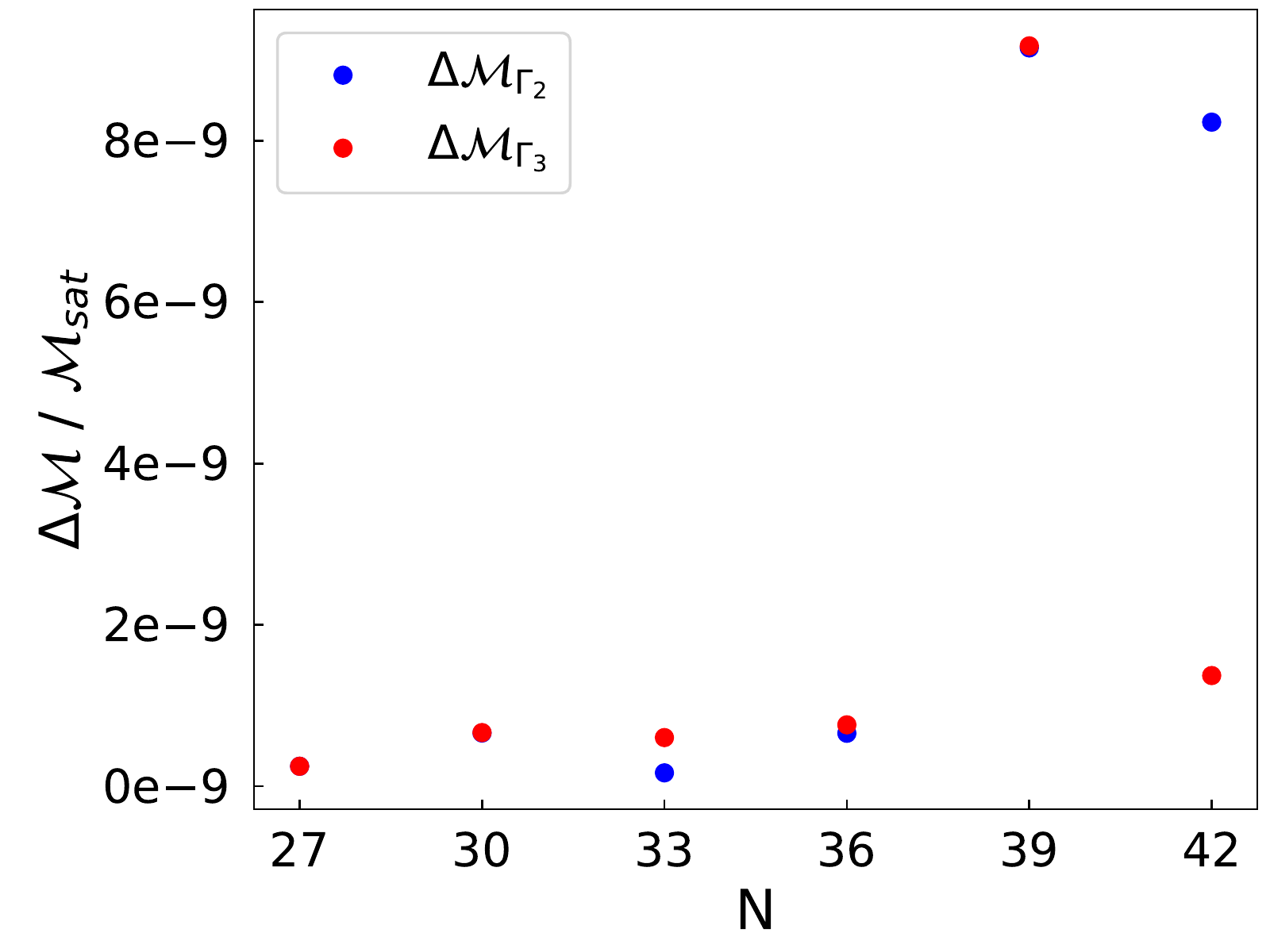}
    \caption{(Color online) Normalized deviations $\Delta \mathcal{M}_{\Gamma_i}(t, h)$ at 
    $t=0.05$, compare \fmref{e-3-1}, 
    for various system sizes. The deviations at $h_-$ (top) are six orders of magnitude 
    bigger than for $h=h_+$ (bottom), see text.}
    \label{f-3-3}
\end{figure} 

As can be seen in \figref{f-3-3} (top), 
the deviations for $h=h_-$ are of the order of $\sfrac{1}{100}$ 
of the saturation magnetization, when only considering subspace $\Hi{}(\Gamma_2(h_-))$. 
When using the greater subspace $\Hi{}(\Gamma_3(h_-))$ the deviations are significantly lower ($\leq 1$~\textperthousand). 
This means, that even at temperatures as low as $t=0.05$ 
there are more than just two subspaces $\Hi{M}$ significantly contributing to the value of $\mathcal{M}(T, h)$ 
at the magnetization jump to the \otp{}, i.e. at $h=h_-$\ .

In \figref{f-3-3} (bottom), the deviations for $h=h_+$ are of the order of $\sim 10^{-9}$,
i.e. six orders of magnitude smaller, even when only regarding  the subspace $\Hi{}(\Gamma_2(h_+))$ 
which consists of only two subspaces $\Hi{M}$. To achieve the same accuracy at $ h=h_-$ 
one would have to consider at least four subspaces $\Hi{M}$.   

\begin{figure}[h!]
    \vspace{0.3cm}
    \centering
    \includegraphics[width=0.9\columnwidth]{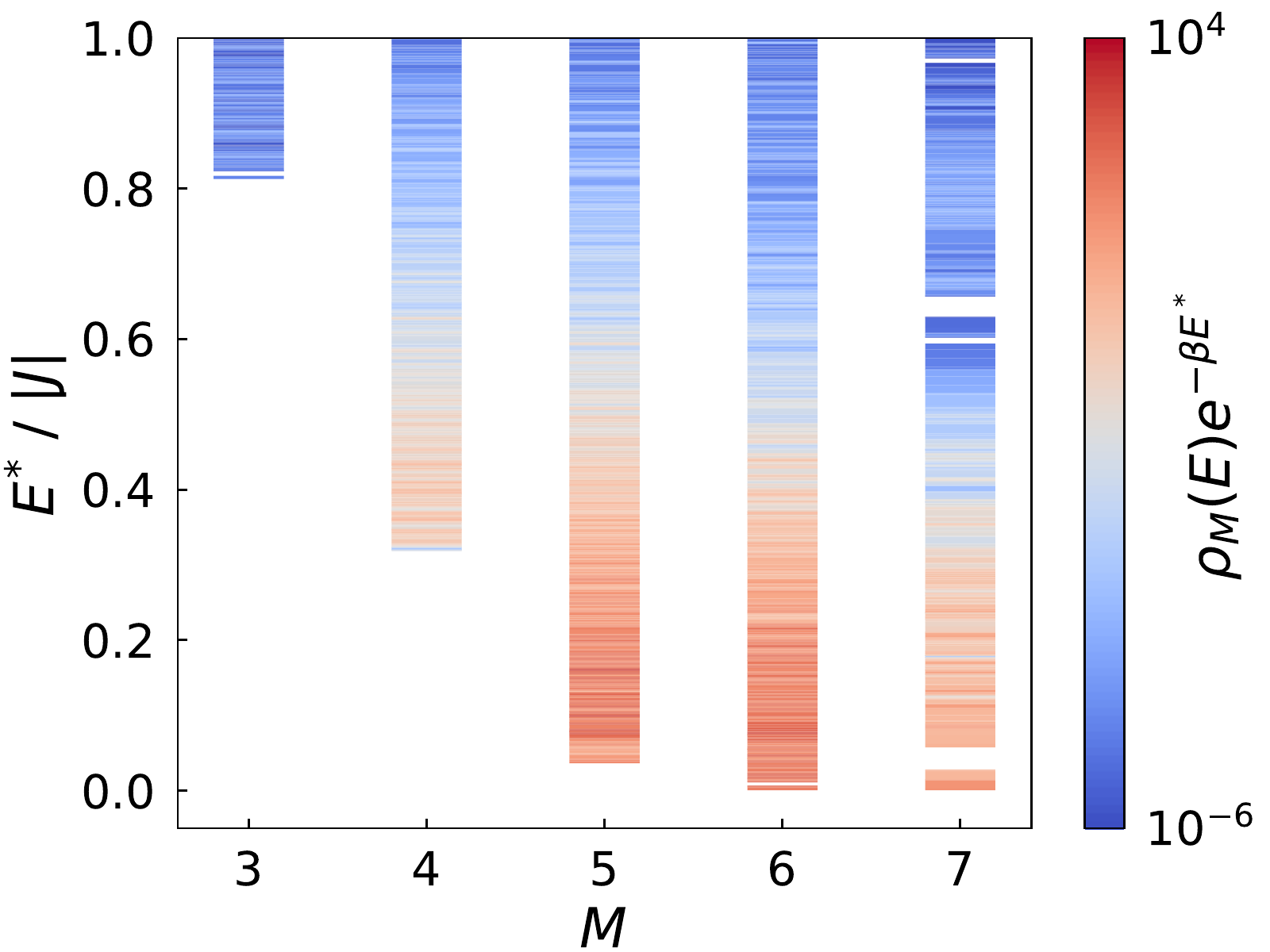}
    
    \includegraphics[width=0.9\columnwidth]{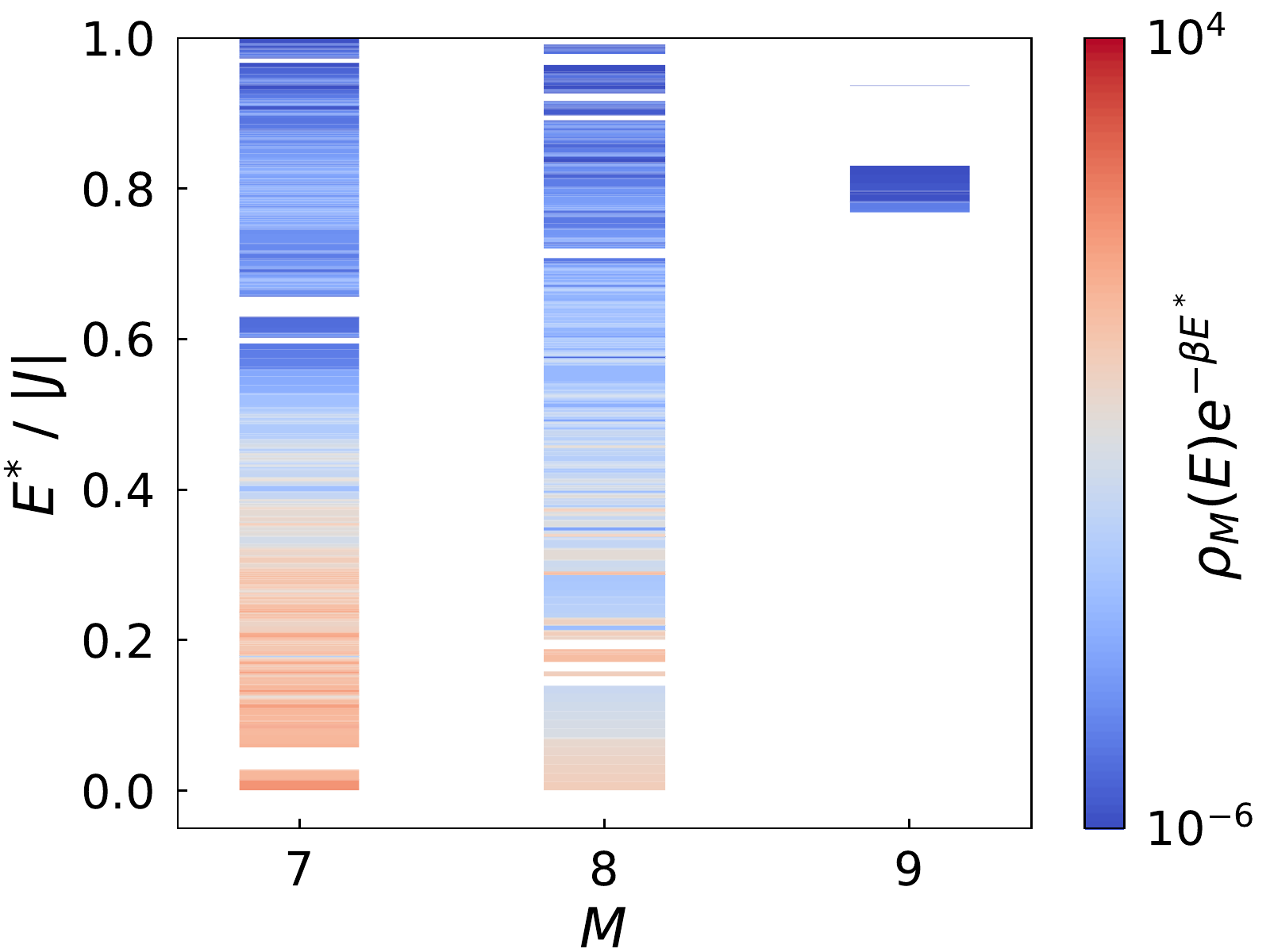}
    \caption{(Color online) Low-energy part of the subspace densities of states $\rho_M(E^*)$ 
    of the energetically lowest subspaces
    for $E^*=(E-E_0(M))/|J| \leq 20 \cdot t$) weighted by the Boltzmann 
    factor at $t=0.05$ and
    calculated at $h_-$ (top) and $h_+$ (bottom) for the KHAF with $N=42$ sites.
    Values smaller than $10^{-7}$ are omitted to improve resolution of the color coding.}
    \label{f-3-4}
\end{figure} 

\paragraph{Density of states:} 
Next, we consider subspace densities of states $\rho_M(E, h)$ at $h_-$ and $h_+$ for $N=42$ sites 
to get an even more profound understanding of the melting process. 
Therefore, we show in \figref{f-3-4} the low-energy part of the densities of states of the energetically lowest subspaces
weighted by the Boltzmann factor at $t=0.05$. Values smaller than $10^{-7}$ are omitted
to improve the resolution of the color coding, which yields the white spaces in \figref{f-3-4} 
although the density of states is not strictly zero.

As can be seen in \figref{f-3-4} (top) the density of low-lying levels is indeed larger in the 
subspace with $M_{1/3}-1$ than in the plateau subspace with $M_{1/3}$, and this density is larger 
than that of the subspace with $M_{1/3}+1$, as was conjectured previously \cite{SSR:PRB18,MMY:PRB20}.
In addition, two subspaces with $M_{1/3}-1$ and $M_{1/3}-2$ contribute to thermal expectation values
at small excitation energies at the 
low-field side of the plateau whereas only one subspace with $M_{1/3}+1$ contributes at the high-field side 
of the plateau.

\begin{figure}[h!]
    \vspace{0.3cm}
    \centering
    \includegraphics[width=0.9\columnwidth]{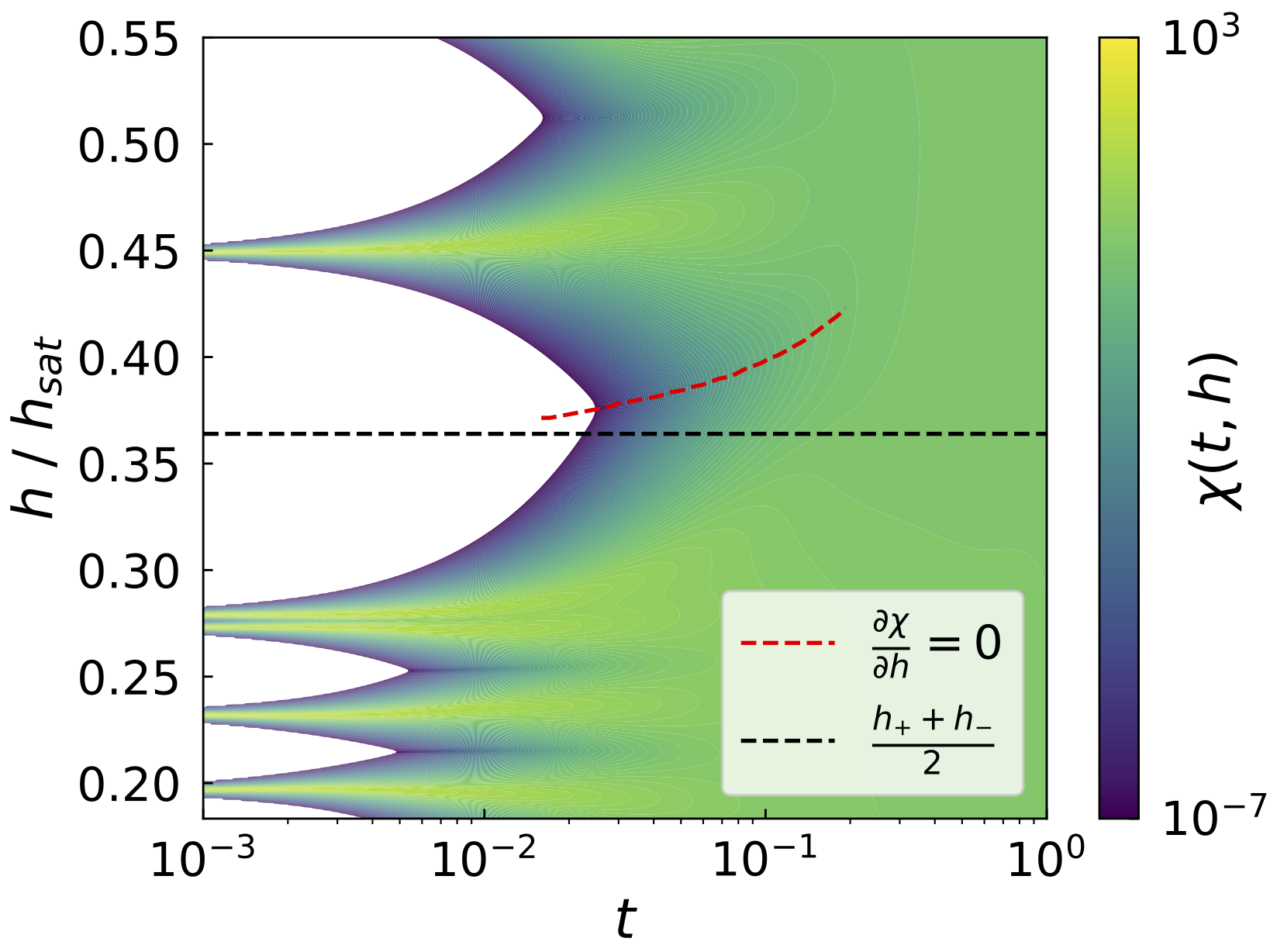}
    \caption{(Color online) The differential magnetic susceptibility $\chi(t,h)$ for the KHAF with $N=42$ sites.
    Values smaller than $10^{-8}$ are omitted to improve resolution of the color coding. 
    The asymmetric melting of the $\sfrac{1}{3}$-plateau is visible as an upturn
    of the region of small susceptibility around $h\sim 0.28\dots 0.45$. 
    It is additionally highlighted
    by the red dashed curve (local minimum of $\chi(t,h)$) that deviates clearly from the symmetric black dashed line.}
    \label{f-3-4-a}
\end{figure} 

Finally, we graphically summarize the melting of magnetization plateaus by plotting the differential 
magnetic susceptibility $\chi(t,h)$ for the KHAF with $N=42$ sites in \figref{f-3-4-a}. 
A flat magnetization plateau corresponds to zero susceptibility, melting increases the susceptibility,
and an asymmetric increase expresses itself as a banana-shaped feature. This behavior is clearly
visible in \figref{f-3-4-a} in the region around $h\sim 0.28\dots 0.45$ and additionally 
highlighted by the red dashed curve (local minimum of $\chi(t,h)$) 
bending towards higher fields compared to a symmetric behavior
shown by the black dashed curve.

\begin{figure}[ht!]
    \centering
    \includegraphics[width=0.9\columnwidth]{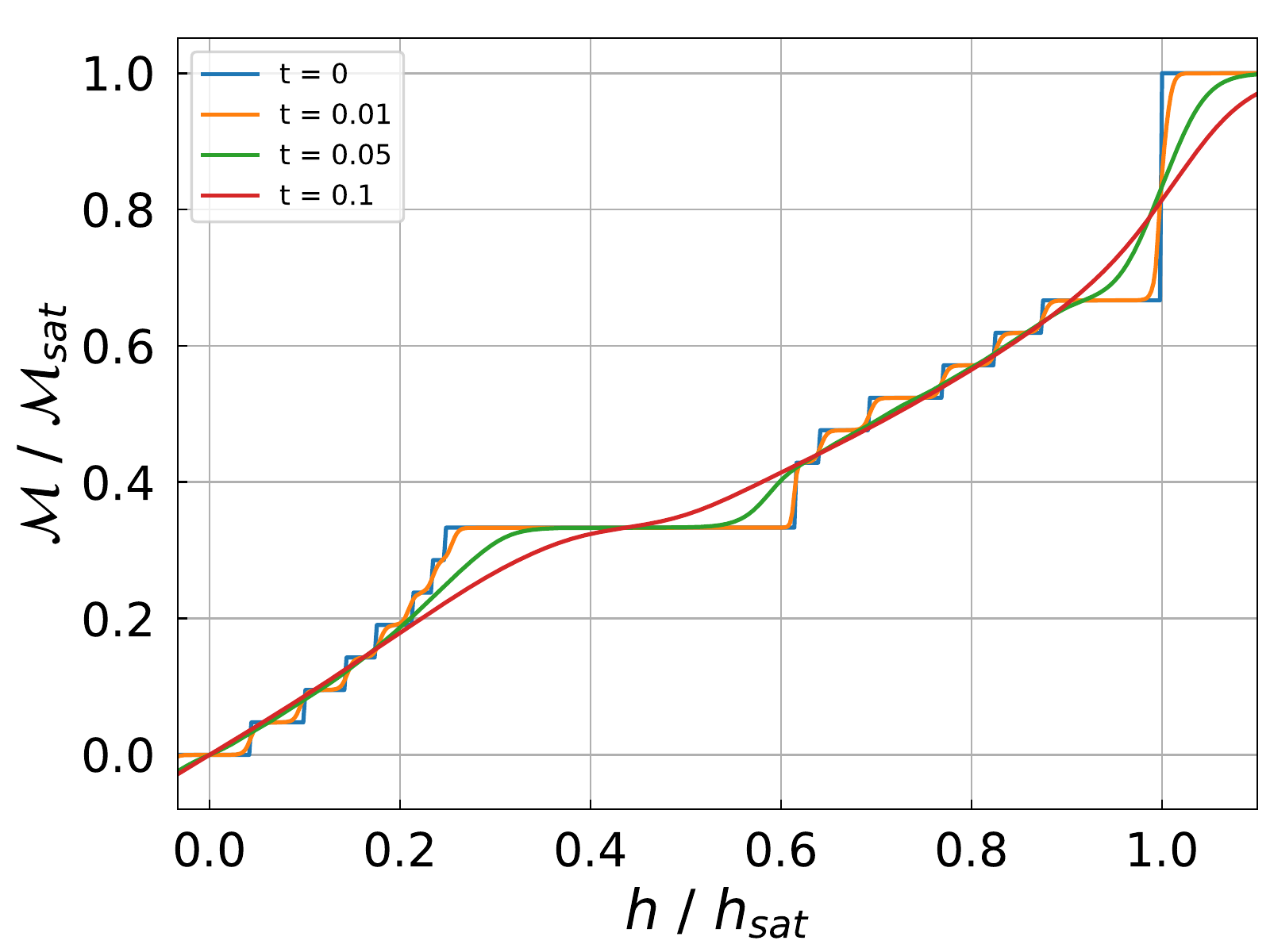}
    \caption{(Color online) Magnetization curve of the square-\kagome lattice with $N=42$ 
    sites at low temperatures 
    in comparison to the zero temperature curve. The $\sfrac{1}{3}$-plateau 
    melts symmetrically.}
    \label{f-3-5}
 \end{figure} 

\begin{figure}[ht!]
    \vspace{0.3cm}
    \centering
    \includegraphics[width=0.9\columnwidth]{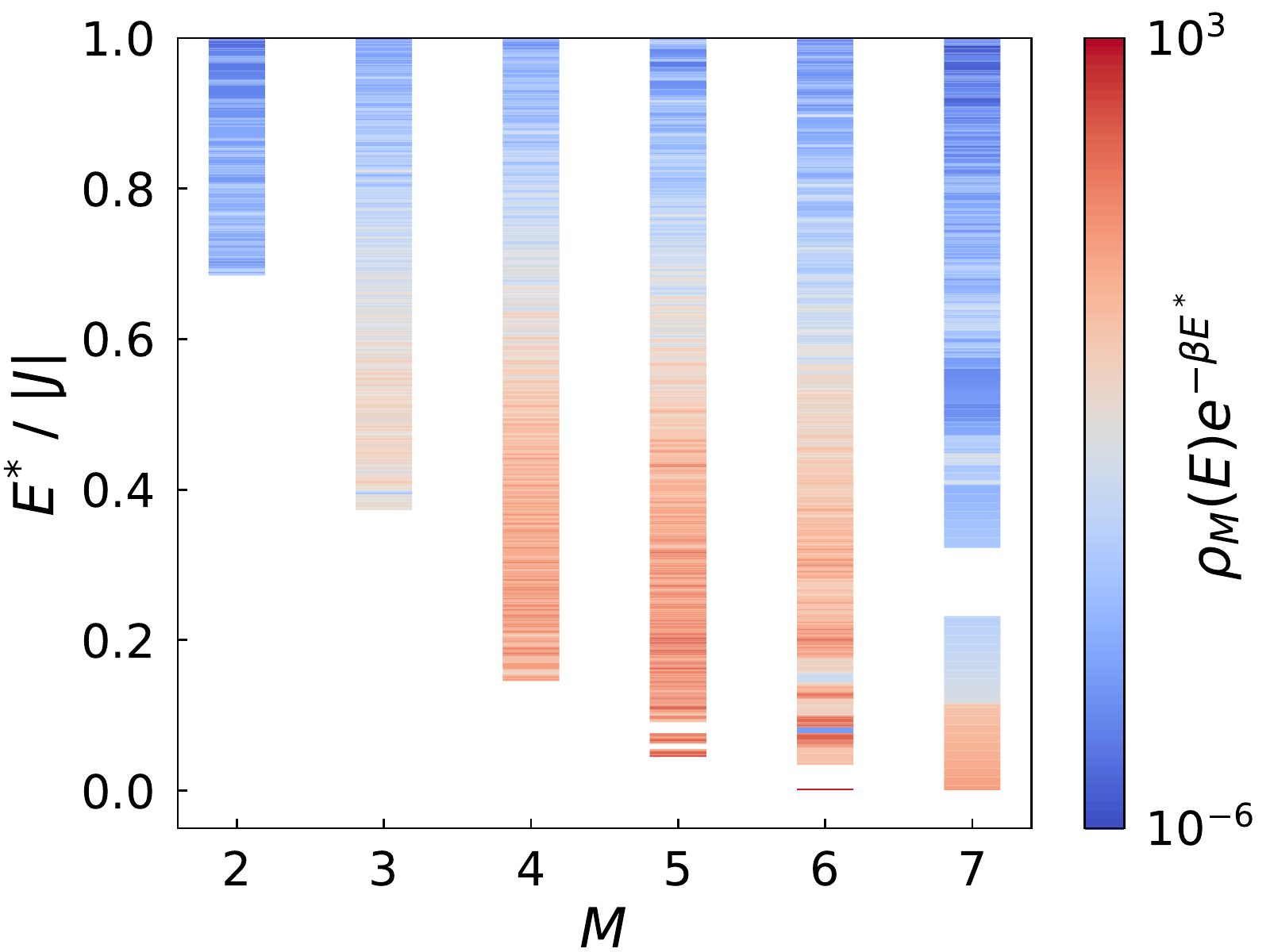}
    
    \includegraphics[width=0.9\columnwidth]{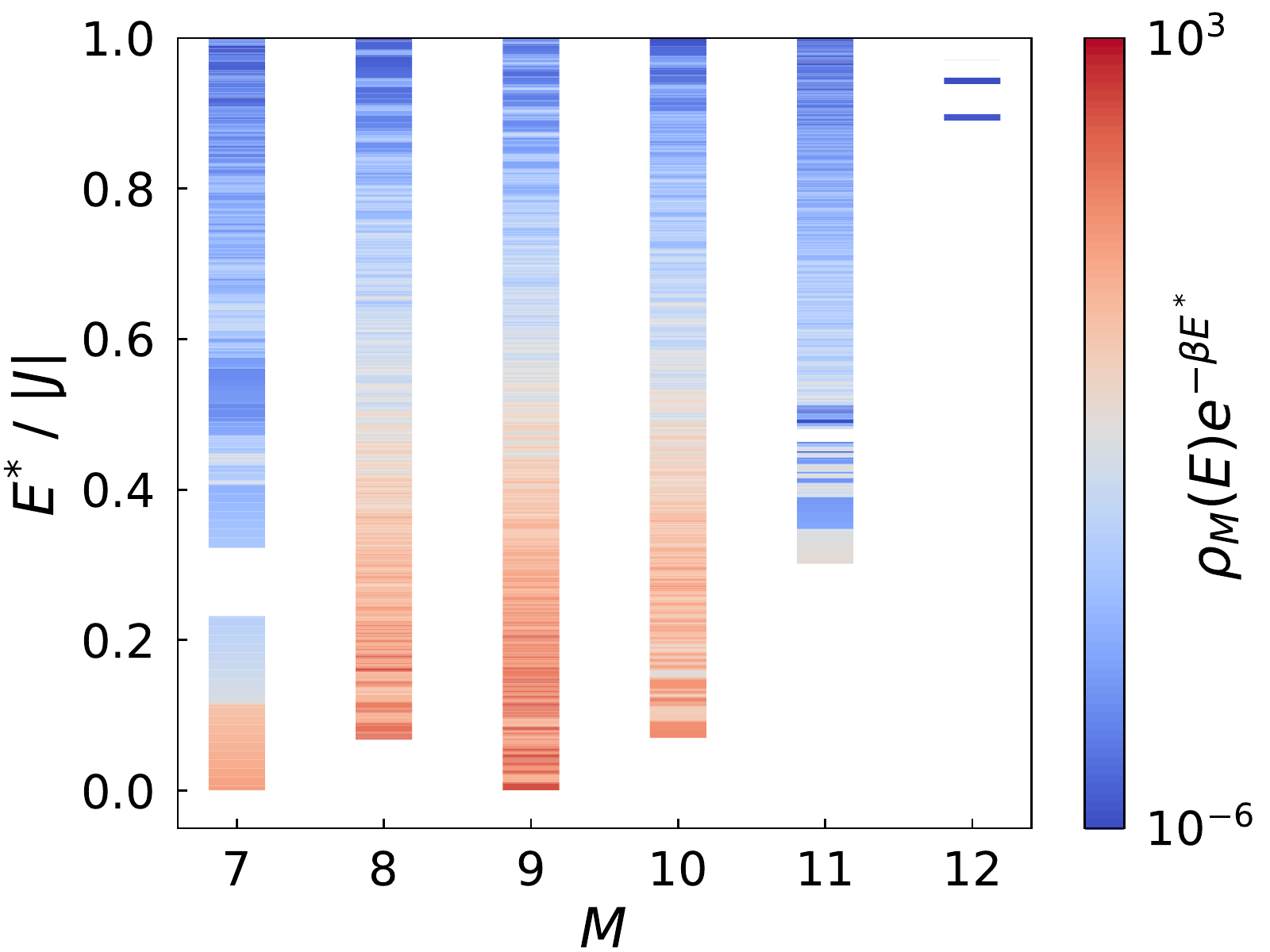}
    \caption{(Color online) Low-energy part of the subspace 
    densities of states $\rho_M(E)$ 
    of the energetically lowest subspaces
    for $E_0(M)/| J | \leq 20 \cdot t$) weighted by the Boltzmann factor at $t=0.05$ and
    calculated at $h_-$ (top) and $h_+$ (bottom) for the SKHAF with $N=42$ sites.
    Values smaller than $10^{-7}$ are omitted to improve resolution of the color coding.}
    \label{f-3-7}
\end{figure} 

\subsection{No asymmetric melting of the SKAHF $\sfrac{1}{3}$-plateau}
\label{sec-3-2}

As a counter example we will consider the 
melting of the $\sfrac{1}{3}$-plateau of 
the square-\kagome lattice antiferromagnet (SKHAF). 
In \figref{f-3-5} one can see that the plateau melts symmetrically 
with increasing temperatures.  

Comparing \figref{f-3-5} of the SKHAF with \figref{f-3-1} of the KHAF
one notices that for the SKHAF the magnetization steps to either 
side of the 
$\sfrac{1}{3}$-plateau have very similar sizes in contrast to our findings 
for the KHAF. 

Looking at the densities of low-lying states of the subspaces with 
$M_{1/3}-1$, $M_{1/3}$, and $M_{1/3}+1$ we find a different trend 
for the SKHAF compared to the KHAF: the density does not steadily 
increase with increasing $M$, it is largest for $M_{1/3}$ and decreases 
when going to either $M_{1/3}-1$ or $M_{1/3}+1$, see \figref{f-3-7}. 
This means roughly that at both edges of the plateau a similar 
number of subspaces contributes significantly to the value 
of the magnetization. 
One should keep in mind, that these contributions consist of the 
density of states multiplied by the Boltzmann factor
as well as the magnetic quantum numbers, and that the 
symmetry we discuss is visible only at rather small temperatures,
i.e., for Boltzmann factors that decrease rapidly with increasing energy.

\begin{figure}[h!]
    \vspace{0.3cm}
    \centering
    \includegraphics[width=0.9\columnwidth]{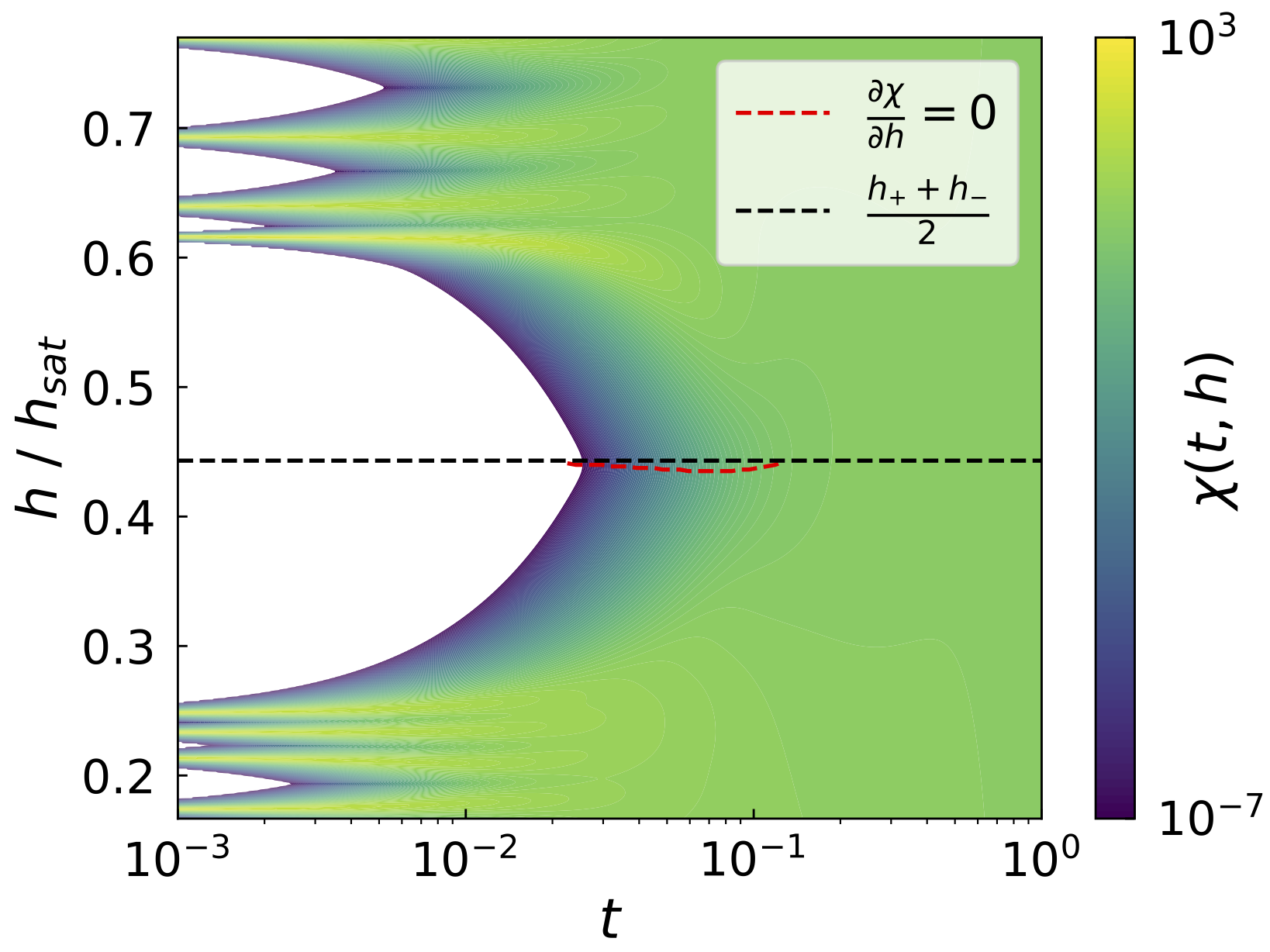}
    \caption{(Color online) The magnetic susceptibility $\chi(h,t)$ for $42$ sites on the square-\kagome lattice. 
    Values smaller than $10^{-8}$ are dropped to improve resolution of the color coding. 
    The symmetric melting of the $\sfrac{1}{3}$-plateau is clearly 
    visible around $h\sim 0.27\dots 0.61$. 
    It is additionally highlighted by the red dashed curve that does not
    deviate from the symmetric black dashed line.}
    \label{f-3-6}
\end{figure} 

We again graphically summarize the melting of magnetization plateaus by plotting the differential 
magnetic susceptibility $\chi(t,h)$ for the SKHAF with $N=42$ sites in \figref{f-3-6}. 
The figure clearly demonstrates that the $\sfrac{1}{3}$-plateau melts symmetrically, 
see region around $h\sim 0.27\dots 0.61$ 
in \figref{f-3-6}. The dashed red curve which marks the minimum of the susceptibility 
does not deviate from the symmetric black dashed line.

\begin{figure}[ht!]
    \vspace{0.3cm}
    \centering
    \includegraphics[width=0.45\columnwidth]{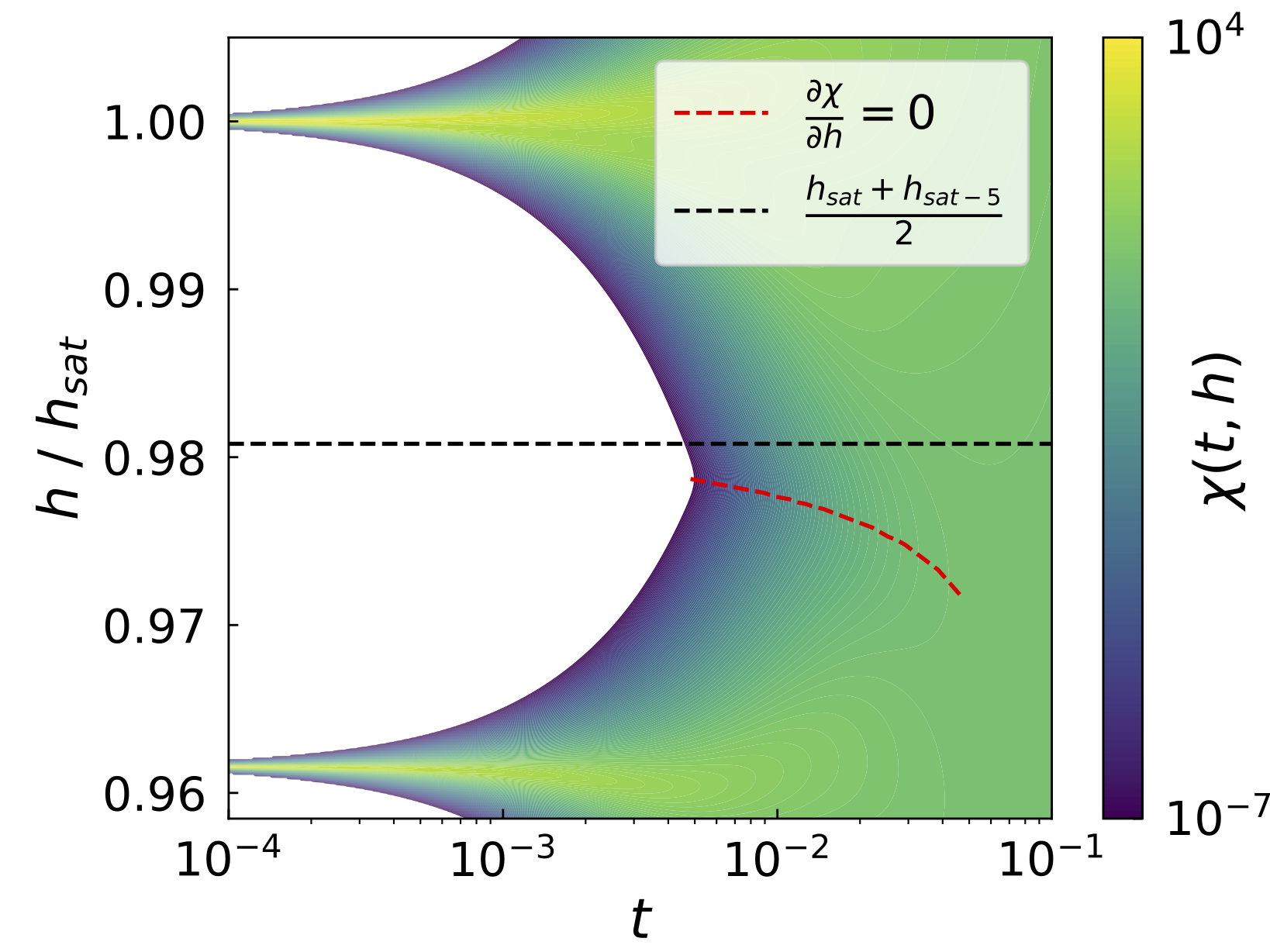}
        \includegraphics[width=0.45\columnwidth]{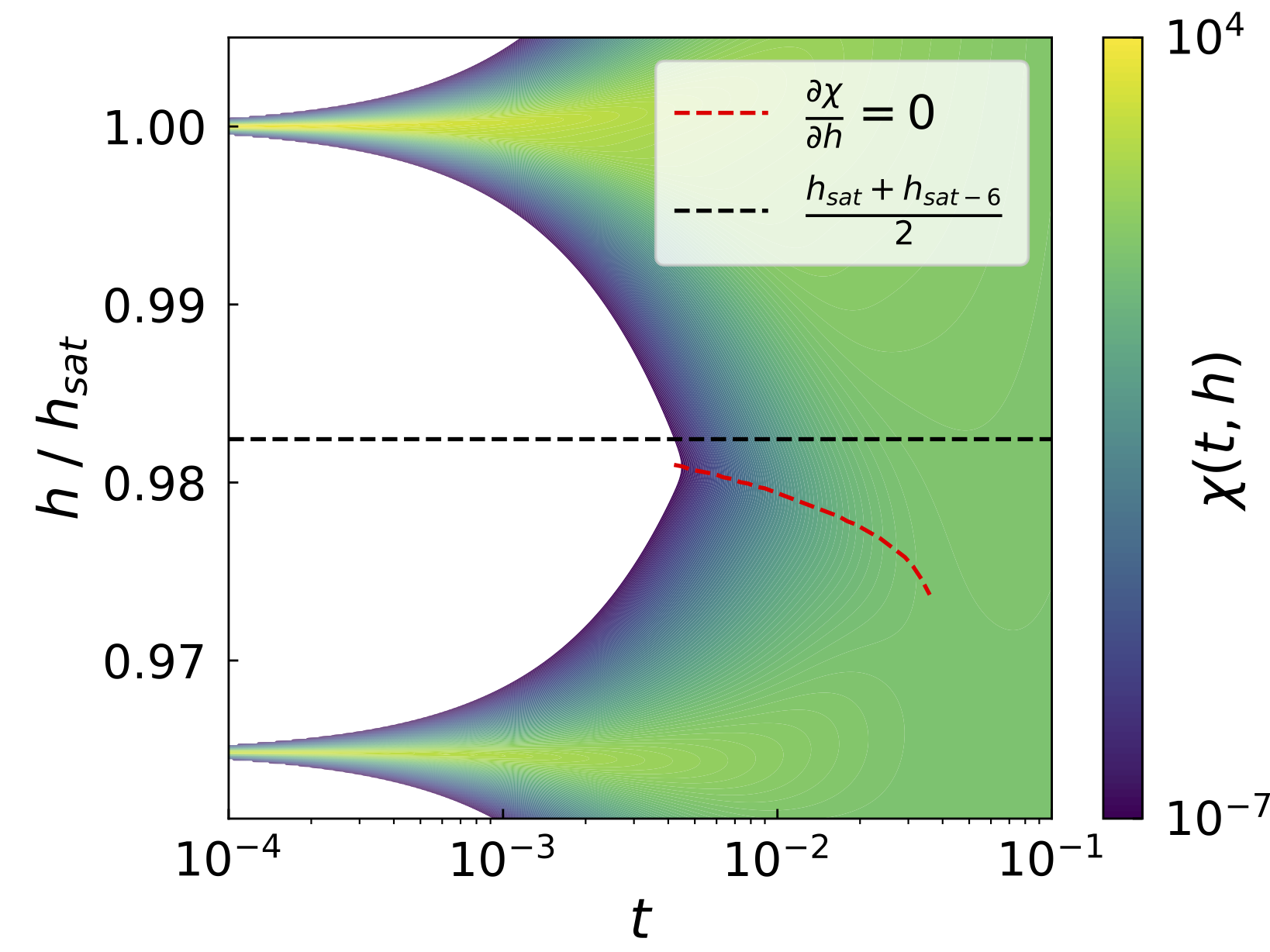}
    
    \includegraphics[width=0.45\columnwidth]{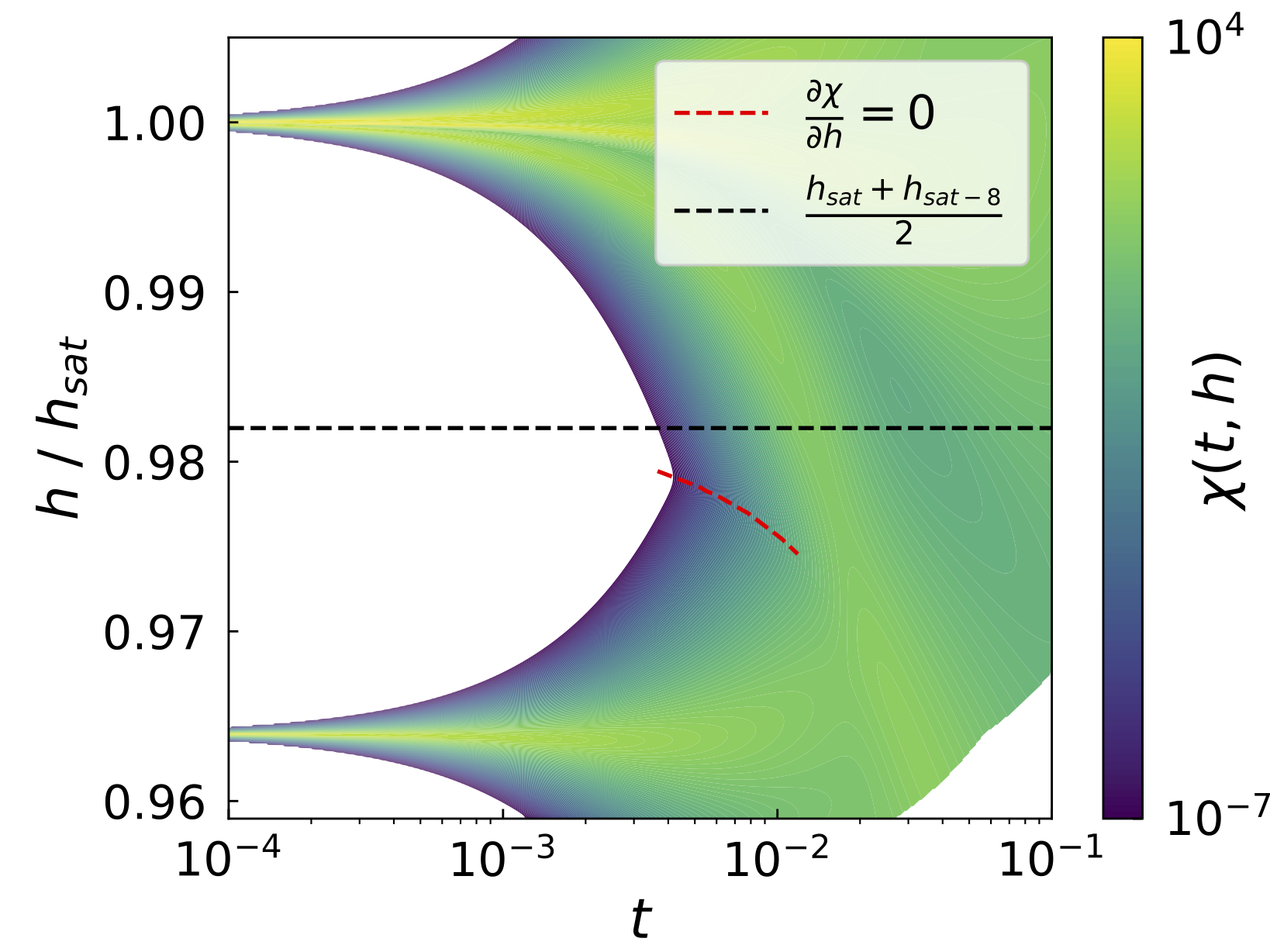}
        \includegraphics[width=0.45\columnwidth]{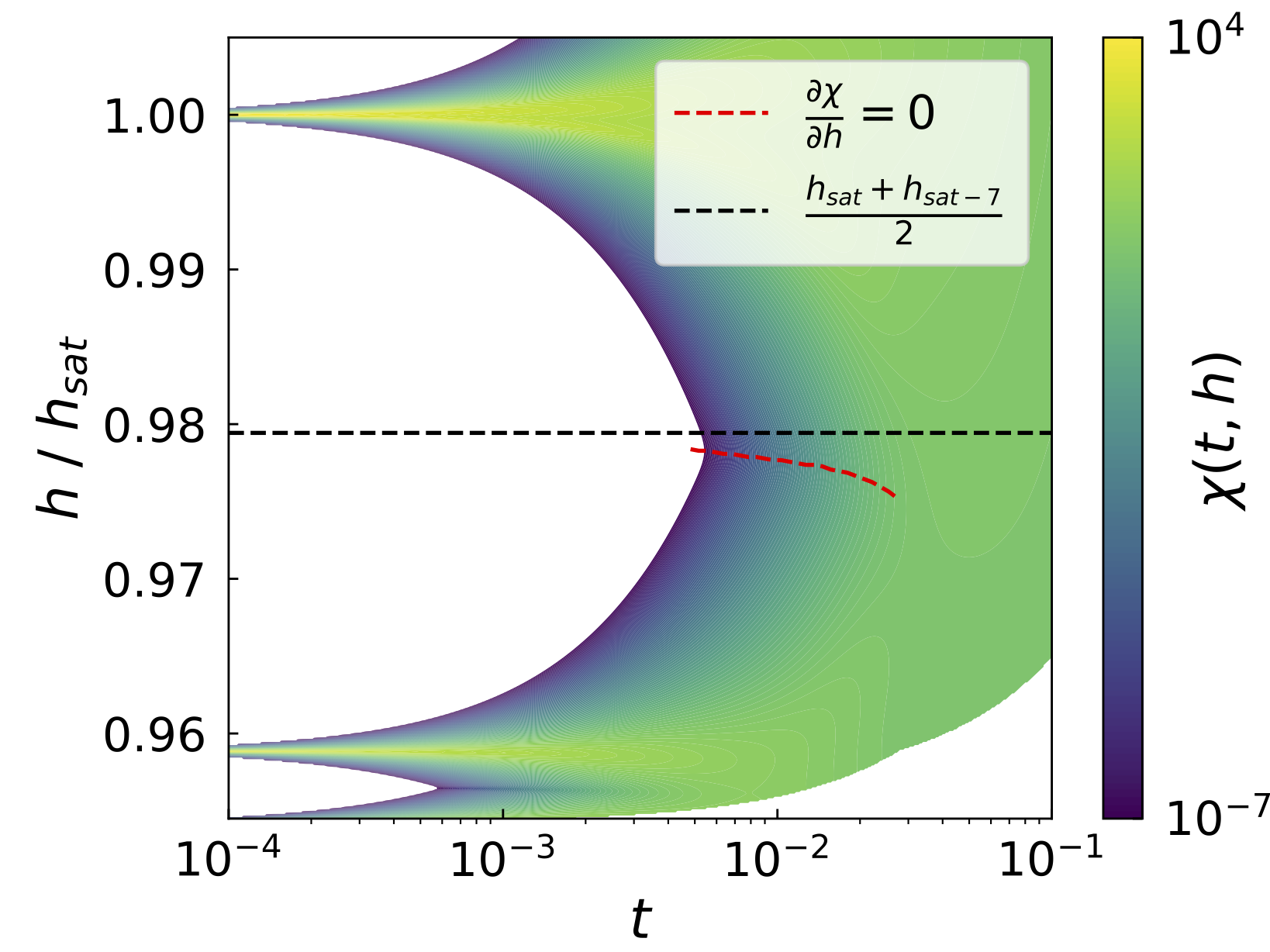}
        \caption{(Color online) Magnetic susceptibility $\chi(h,t)$ for the KHAF of $N=45, 54, 63, 72$ 
        sites (clockwise from top left).
    Values smaller than $10^{-8}$ are dropped to improve the resolution of the color coding. 
    The $\sfrac{7}{9}$-plateau melts asymmetrically with a downturn of the minimum of the susceptibility, 
    see red dashed curves.}
    \label{f-3-9}
\end{figure} 

\begin{figure}[ht!]
    \vspace{0.3cm}
    \centering
    \includegraphics[width=0.45\columnwidth]{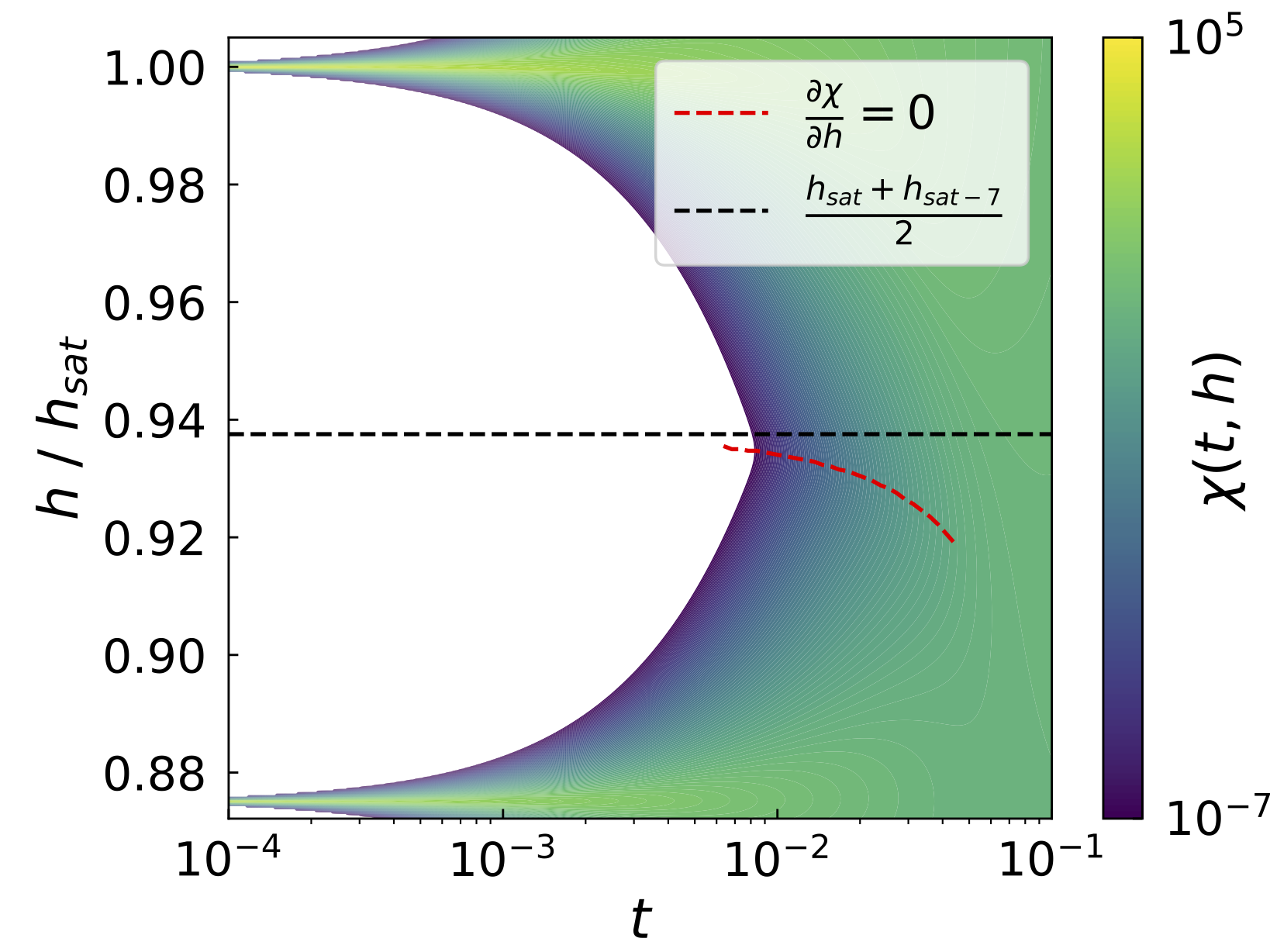}
        \includegraphics[width=0.45\columnwidth]{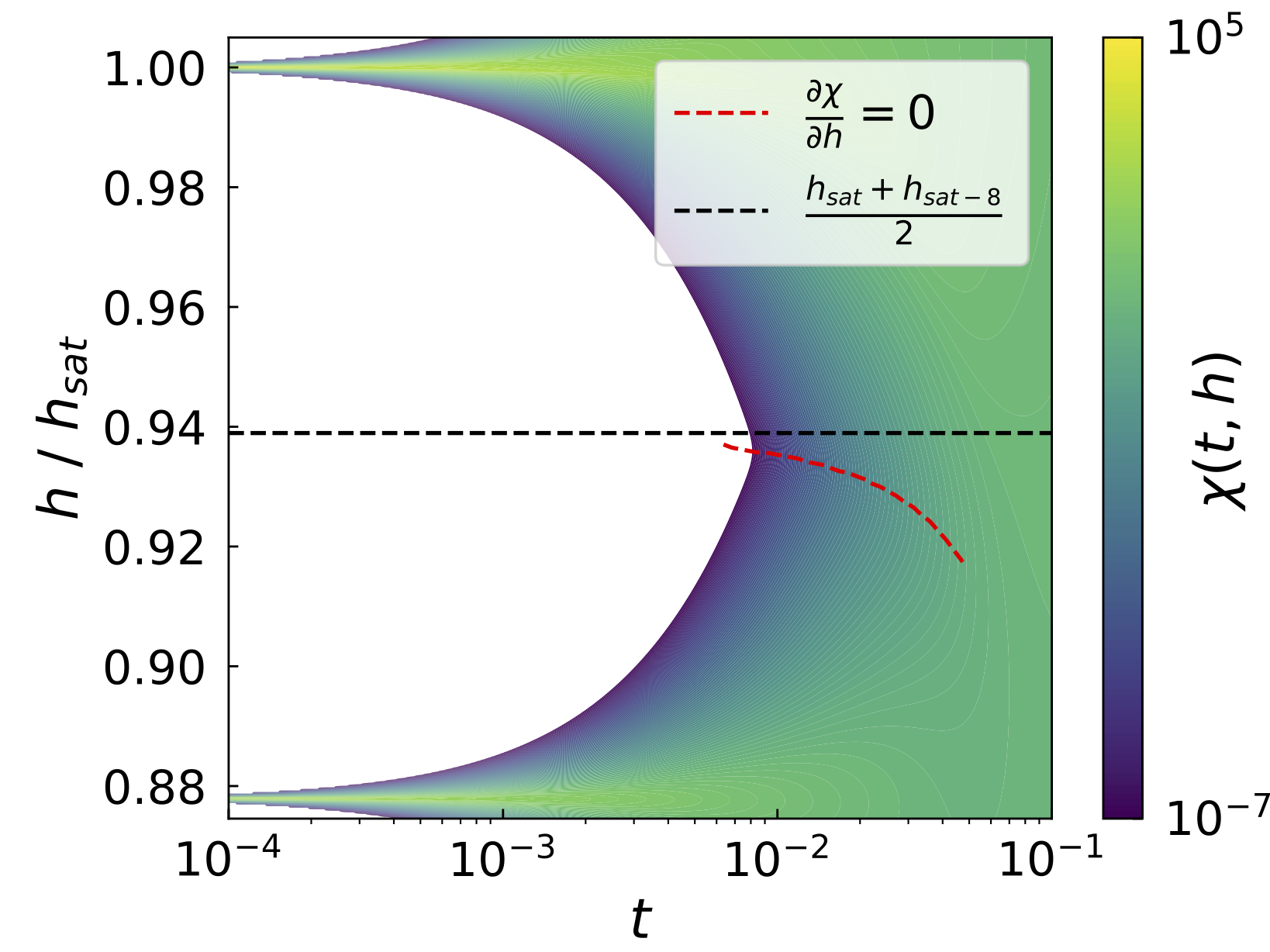}
        
    \includegraphics[width=0.45\columnwidth]{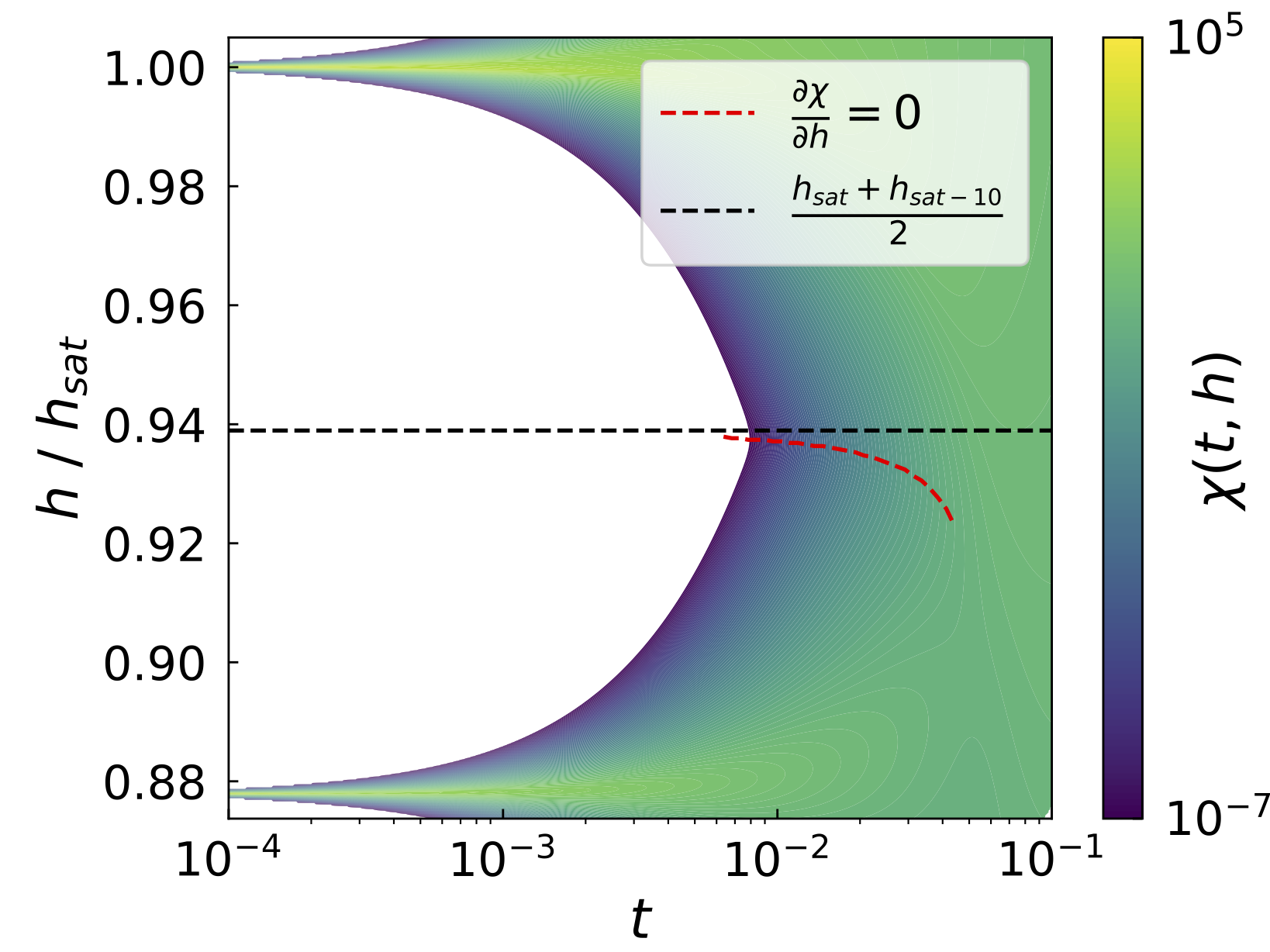}
        \includegraphics[width=0.45\columnwidth]{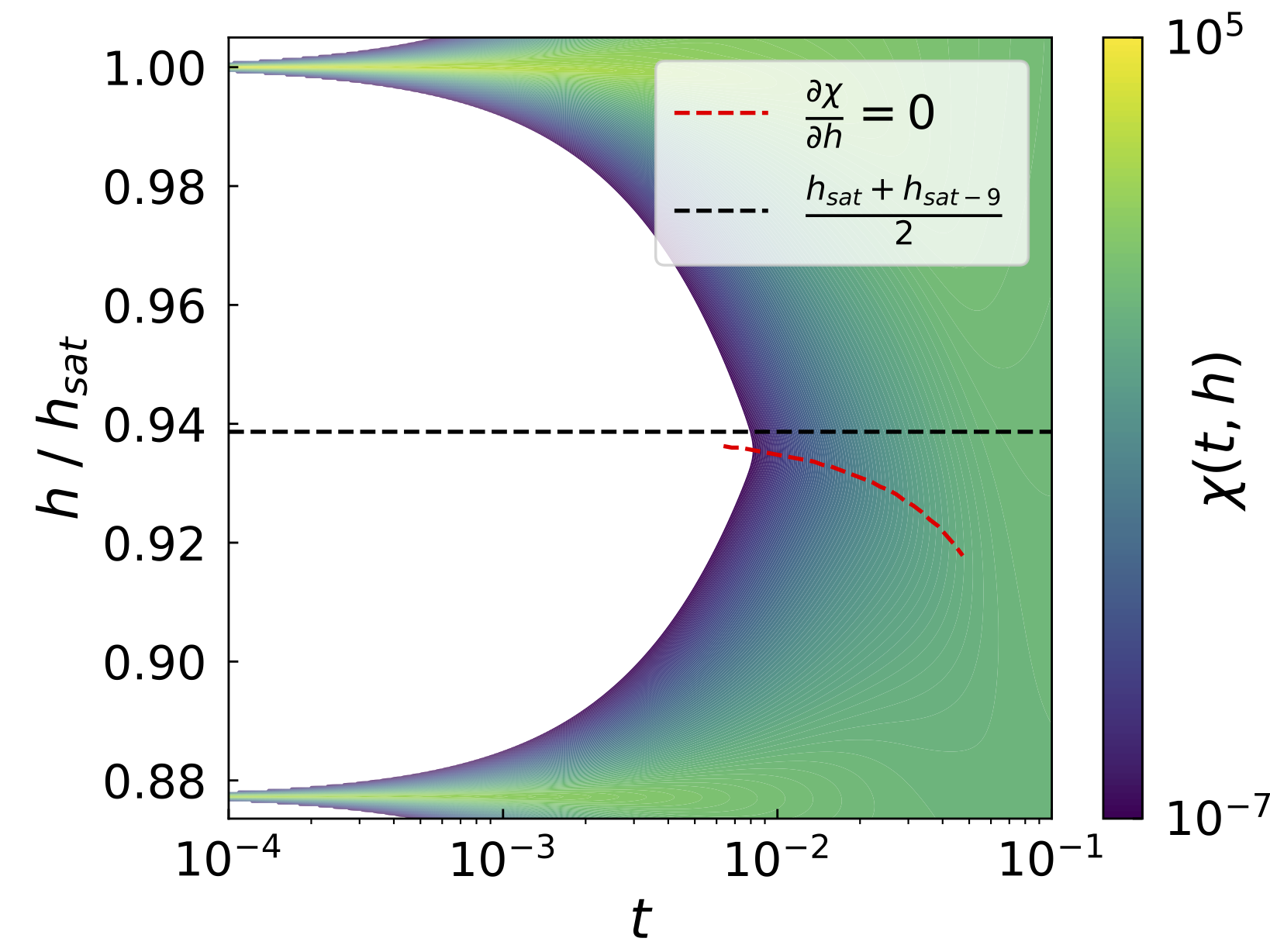}
        \caption{(Color online) Magnetic susceptibility $\chi(h,t)$ for the SKHAF 
        of $N=42, 48, 54, 60$ 
        sites (clockwise from top left).
    Values smaller than $10^{-8}$ are dropped to improve the resolution of the color coding. 
    The $\sfrac{2}{3}$-plateau melts asymmetrically with a downturn of the minimum of the susceptibility, 
    see red dashed curves.}
    \label{f-3-10}
\end{figure} 

\subsection{Excursus -- plateau next to saturation}
\label{sec-3-3}

Although the $\sfrac{1}{3}$-plateau melts differently for the KHAF and the SKHAF
the magnetization plateau that precedes the magnetization jump to saturation, compare
\cite{SHS:PRL02,SSR:PRB18,RDS:PRB22}, melts asymmetrically for both lattices as depicted 
by the dashed red curves of minimal susceptibility in
Figs.~\xref{f-3-9} and \xref{f-3-10}. 
In contrast to the asymmetric melting of the $\sfrac{1}{3}$-plateau of the KHAF 
here we observe a pronounced bending towards lower fields.
This feature is related to the very existence of a flat one-magnon band 
\cite{DRM:IJMPB15,LAF:APX18} 
and is therefore a generic effect of flat-band quantum magnets. 
For spin systems with a flat one-magnon band the structure of the density of
low-lying states at and below the saturation field is very similar 
and dominated by localized multi-magnon states that are degenerate 
at the saturation field and split up for smaller fields as well as by the nearly exponentially 
growing dimension of subspaces $\Hi{M}$ with decreasing $|M|$.
Therefore, we may expect that the asymmetric melting of this high-field plateau is
present also if the flat-band becomes slightly dispersive, as for example in
the diamond-shaped compound azurite \cite{KFC:PRL05,KFC:PTPS05,JOK:PRL11}.

A possible difference between various flat-band systems could be given by the magnetization 
of the plateau that precedes the magnetization jump to saturation.
For the KHAF this is $\sfrac{7}{9}$ and for the SKHAF 
this is $\sfrac{2}{3}$ of the saturation magnetization, respectively 
\cite{SHS:PRL02,SSR:PRB18,RDS:PRB22}. Figure~\xref{f-3-9} shows the magnetic susceptibility of the KHAF
for fields close to saturation and temperatures $t=10^{-4}, \dots 10^{-1}$ for 
$N=45, 54, 63, 72$ sites (clockwise from top left). 
Figure~\xref{f-3-10} displays the magnetic susceptibility of the SKHAF
for fields close to saturation and temperatures $t=10^{-4}, \dots 10^{-1}$
for $N=42, 48, 54, 60$ sites (clockwise from top left). 
Here $h_{\text{sat}}$ denotes the saturation field, and $h_{\text{sat}-k}$
denotes the low-field end of the plateau, where $k$ is the largest number
of localized multi-magnon states that fits on the respective size of
the lattice.

All cases clearly exhibit asymmetric melting towards lower fields 
for increasing temperatures.


\section{Discussion and conclusions}
\label{sec-4}

The experimental magnetization is often not directly determined but via
its derivative with respect to the applied field, i.e., the susceptibility,
in particular for instance in pulsed field measurements, see, e.g. 
\cite{KFC:PTPS05,ONO:NC19,FMM:NC20} for recent related examples.
Therefore, the question how plateaus deform with
elevated temperatures is very relevant for the interpretation of measurements 
of the magnetization. Asymmetric melting means that the minimum of the susceptibility
moves away from the center of the $(T=0)$-plateau with rising temperature.

Looking at our findings, we tend to conclude that the main cause of the asymmetric melting 
of the $\sfrac{1}{3}$-plateau of the KHAF 
is that only two subspaces contribute to the magnetization at the right edge of the $\sfrac{1}{3}$-plateau
at low temperatures whereas at the left edge several subspaces with a broader spread of magnetic quantum numbers 
contribute for the same temperature. 
In the case of the SKHAF the latter is the case at both ends of the plateau 
which results in a more symmetric melting.

\begin{acknowledgments}
This work was supported by the Deutsche Forschungsgemeinschaft (DFG RI 615/25-1 and SCHN 615/28-1). 
Supercomputing time at the Leibniz Center in Garching (pr62to) is gratefully acknowledged.
\end{acknowledgments}


\begin{thebibliography}{10}

\bibitem{LSM:PRB19}
A.~M. L\"auchli, J.~Sudan, and R.~Moessner: Phys. Rev. B {\bfseries 100} (2019)
  155142.

\bibitem{Yan2011}
S.~Yan, D.~A. Huse, and S.~R. White: Science {\bfseries 332} (2011) 1173.

\bibitem{IBP:PRB11}
Y.~Iqbal, F.~Becca, and D.~Poilblanc: Phys. Rev. B {\bfseries 84} (2011)
  020407.

\bibitem{DMS:PRL12}
S.~Depenbrock, I.~P. McCulloch, and U.~Schollw\"ock: Phys. Rev. Lett.
  {\bfseries 109} (2012) 067201.

\bibitem{Laeuchli2011}
A.~M. L\"auchli, J.~Sudan, and E.~S. S{\o}rensen: Phys. Rev. B {\bfseries 83}
  (2011) 212401.

\bibitem{Iqbal2013}
Y.~Iqbal, F.~Becca, S.~Sorella, and D.~Poilblanc: Phys. Rev. B {\bfseries 87}
  (2013) 060405.

\bibitem{Nor:RMP16}
M.~R. Norman: Rev. Mod. Phys. {\bfseries 88} (2016) 041002.

\bibitem{Pollmann2017}
Y.~He, M.~P. Zaletel, M.Oshikawa, and F.~Pollmann: Phys. Rev. X {\bfseries 7}
  (2017) 031020.

\bibitem{Xie2017}
H.~J. Liao, Z.~Y. Xie, J.~Chen, Z.~Y. Liu, H.~D. Xie, R.~Z. Huang, B.~Normand,
  and T.~Xiang: Phys. Rev. Lett. {\bfseries 118} (2017) 137202.

\bibitem{elstner1994}
N.~Elstner and A.~P. Young: Phys. Rev. B {\bfseries 50} (1994) 6871.

\bibitem{NaM:PRB95}
T.~Nakamura and S.~Miyashita: Phys. Rev. B {\bfseries 52} (1995) 9174.

\bibitem{ToR:PRB96}
P.~Tomczak and J.~Richter: Phys. Rev. B {\bfseries 54} (1996) 9004.

\bibitem{WEB:EPJB98}
C.~Waldtmann, H.~U. Everts, B.~Bernu, C.~Lhuillier, P.~Sindzingre,
  P.~Lecheminant, and L.~Pierre: Eur. Phys. J. B {\bfseries 2} (1998) 501.

\bibitem{Lhuillier_thermo_PRL2000}
P.~Sindzingre, G.~Misguich, C.~Lhuillier, B.~Bernu, L.~Pierce, C.~Waldtmann,
  and H.~U. Everts: Phys. Rev. Lett. {\bfseries 84} (2000) 2953.

\bibitem{Yu2000}
W.~Yu and S.~Feng: EPJB {\bfseries 13} (2000) 265.

\bibitem{Bernhard2002}
B.~H. Bernhard, B.~Canals, and C.~Lacroix: Phys. Rev. B {\bfseries 66} (2002)
  104424.

\bibitem{Bernu2005}
G.~Misguich and B.~Bernu: Phys. Rev. B {\bfseries 71} (2005) 014417.

\bibitem{MiS:EPJB07}
G.~Misguich and P.~Sindzingre: Eur. Phys. J B {\bfseries 59} (2007) 305.

\bibitem{RBS:PRE07}
M.~Rigol, T.~Bryant, and R.~R.~P. Singh: Phys. Rev. E {\bfseries 75} (2007)
  061118.

\bibitem{Lohmann2014}
A.~Lohmann, H.-J. Schmidt, and J.~Richter: Phys. Rev. B {\bfseries 89} (2014)
  014415.

\bibitem{Shimokawa2016}
T.~Shimokawa and H.~Kawamura: J. Phys. Soc. Jpn. {\bfseries 85} (2016) 113702.

\bibitem{ShS:PRB18}
N.~E. Sherman and R.~R.~P. Singh: Phys. Rev. B {\bfseries 97} (2018) 014423.

\bibitem{MZR:PRB18}
P.~M\"uller, A.~Zander, and J.~Richter: Phys. Rev. B {\bfseries 98} (2018)
  024414.

\bibitem{CRL:SB18}
X.~Chen, S.-J. Ran, T.~Liu, C.~Peng, Y.-Z. Huang, and G.~Su: Science Bulletin
  {\bfseries 63} (2018) 1545 .

\bibitem{Hida2001}
K.~Hida: J. Phys. Soc. Jpn. {\bfseries 70} (2001) 3673.

\bibitem{SHS:PRL02}
J.~Schulenburg, A.~Honecker, J.~Schnack, J.~Richter, and H.-J. Schmidt: Phys.
  Rev. Lett. {\bfseries 88} (2002) 167207.

\bibitem{HSR:JP04}
A.~Honecker, J.~Schulenburg, and J.~Richter: J. Phys.: Condens. Matter
  {\bfseries 16} (2004) S749.

\bibitem{derzhko2004finite}
O.~Derzhko and J.~Richter: Physical Review B {\bfseries 70} (2004) 104415.

\bibitem{zhitomirsky2004exact}
M.~E. Zhitomirsky and H.~Tsunetsugu: Phys. Rev. B {\bfseries 70} (2004) 100403.

\bibitem{CGH:PRB05}
D.~C. Cabra, M.~D. Grynberg, P.~C.~W. Holdsworth, A.~Honecker, P.~Pujol,
  J.~Richter, D.~Schmalfu\ss{}, and J.~Schulenburg: Phys. Rev. B {\bfseries 71}
  (2005) 144420.

\bibitem{SaN:PRB11}
T.~Sakai and H.~Nakano: Phys. Rev. B {\bfseries 83} (2011) 100405.

\bibitem{GMM:JPCM11}
M.~V. Gvozdikova, P.-E. Melchy, and M.~E. Zhitomirsky: J.Phys. Condens. Matter
  {\bfseries 23} (2011) 164209.

\bibitem{Nishimoto2013}
S.~Nishimoto, N.~Shibata, and C.~Hotta: Nature Com. {\bfseries 4} (2013) 2287.

\bibitem{CDH:PRB13}
S.~Capponi, O.~Derzhko, A.~Honecker, A.~M. L\"auchli, and J.~Richter: Phys.
  Rev. B {\bfseries 88} (2013) 144416.

\bibitem{NaS:JPSJ14}
H.~Nakano and T.~Sakai: J. Phys. Soc. Jpn. {\bfseries 83} (2014) 104710.

\bibitem{KPO:PRB16}
A.~Kshetrimayum, T.~Picot, R.~Or\'us, and D.~Poilblanc: Phys. Rev. B {\bfseries
  94} (2016) 235146.

\bibitem{PMH:PRB18}
X.~Plat, T.~Momoi, and C.~Hotta: Phys. Rev. B {\bfseries 98} (2018) 014415.

\bibitem{NaS:JPSJ18}
H.~Nakano and T.~Sakai: J. Phys. Soc. Jpn. {\bfseries 87} (2018) 063706.

\bibitem{HCG:PB05}
A.~Honecker, D.~C. Cabra, M.~D. Grynberg, P.~C.~W. Holdsworth, P.~Pujol,
  J.~Richter, D.~Schmalfuss, and J.~Schulenburg: Physica B {\bfseries 359}
  (2005) 1391.

\bibitem{DRH:LTP07}
O.~Derzhko, J.~Richter, A.~Honecker, and H.-J. Schmidt: Low Temp. Phys.
  {\bfseries 33} (2007) 745.

\bibitem{GeS:PRB22}
M.~Gen and H.~Suwa: Phys. Rev. B {\bfseries 105} (2022) 174424.

\bibitem{Yos:JPSJ22}
H.~K. Yoshida: J. Phys. Soc. Jpn. {\bfseries 91} (2022) 101003.

\bibitem{SSR:PRB18}
J.~Schnack, J.~Schulenburg, and J.~Richter: Phys. Rev. B {\bfseries 98} (2018)
  094423.

\bibitem{MMY:PRB20}
T.~Misawa, Y.~Motoyama, and Y.~Yamaji: Phys. Rev. B {\bfseries 102} (2020)
  094419.

\bibitem{SaN:JKPS13}
T.~Sakai and H.~Nakano: J. Kor. Phys. Soc. {\bfseries 63} (2013) 601.

\bibitem{JaP:PRB94}
J.~Jakli\ifmmode~\check{c}\else \v{c}\fi{} and P.~Prelov\ifmmode~\check{s}\else
  \v{s}\fi{}ek: Phys. Rev. B {\bfseries 49} (1994) 5065.

\bibitem{HaD:PRE00}
A.~Hams and H.~De~Raedt: Phys. Rev. E {\bfseries 62} (2000) 4365.

\bibitem{ADE:PRB03}
M.~Aichhorn, M.~Daghofer, H.~G. Evertz, and W.~von~der Linden: Phys. Rev. B
  {\bfseries 67} (2003) 161103(R).

\bibitem{ScW:EPJB10}
J.~Schnack and O.~Wendland: Eur. Phys. J. B {\bfseries 78} (2010) 535.

\bibitem{SuS:PRL12}
S.~Sugiura and A.~Shimizu: Phys. Rev. Lett. {\bfseries 108} (2012) 240401.

\bibitem{SuS:PRL13}
S.~Sugiura and A.~Shimizu: Phys. Rev. Lett. {\bfseries 111} (2013) 010401.

\bibitem{ScT:PR17}
B.~Schmidt and P.~Thalmeier: Phys. Rep. {\bfseries 703} (2017) 1 .

\bibitem{PrK:PRB18}
P.~Prelov\ifmmode~\check{s}\else \v{s}\fi{}ek and J.~Kokalj: Phys. Rev. B
  {\bfseries 98} (2018) 035107.

\bibitem{OAD:PRE18}
S.~Okamoto, G.~Alvarez, E.~Dagotto, and T.~Tohyama: Phys. Rev. E {\bfseries 97}
  (2018) 043308.

\bibitem{IMN:IEEE19}
K.~Inoue, Y.~Maeda, H.~Nakano, and Y.~Fukumoto: IEEE Transactions on Magnetics
  {\bfseries 55} (2019) 1.

\bibitem{MoT:PRR20}
K.~Morita and T.~Tohyama: Phys. Rev. Research {\bfseries 2} (2020) 013205.

\bibitem{RST:CMP09}
J.~Richter, J.~Schulenburg, P.~Tomczak, and D.~Schmalfu{\ss}: Cond. Matter
  Phys. {\bfseries 12} (2009) 507.

\bibitem{NaS:JPSJ13}
H.~Nakano and T.~Sakai: J. Phys. Soc. Jpn. {\bfseries 82} (2013) 083709.

\bibitem{RDS:PRB22}
J.~Richter, O.~Derzhko, and J.~Schnack: Phys. Rev. B {\bfseries 105} (2022)
  144427.

\bibitem{spin:258}
J.~Schulenburg.
\newblock {\em spinpack 2.58}.
\newblock Magdeburg University, 2019.

\bibitem{SRS:PRR20}
J.~Schnack, J.~Richter, and R.~Steinigeweg: Phys. Rev. Research {\bfseries 2}
  (2020) 013186.

\bibitem{DRM:IJMPB15}
O.~Derzhko, J.~Richter, and M.~Maksymenko: Int. J. Mod. Phys. B {\bfseries 29}
  (2015) 1530007.

\bibitem{LAF:APX18}
D.~Leykam, A.~Andreanov, and S.~Flach: Advances in Physics: X {\bfseries 3}
  (2018) 1473052.

\bibitem{KFC:PRL05}
H.~Kikuchi, Y.~Fujii, M.~Chiba, S.~Mitsudo, T.~Idehara, T.~Tonegawa,
  K.~Okamoto, T.~Sakai, T.~Kuwai, and H.~Ohta: Phys. Rev. Lett. {\bfseries 94}
  (2005) 227201.

\bibitem{KFC:PTPS05}
H.~Kikuchi, Y.~Fujii, M.~Chiba, S.~Mitsudo, T.~Idehara, T.~Tonegawa,
  K.~Okamoto, T.~Sakai, T.~Kuwai, K.~Kindo, A.~Matsuo, W.~Higemoto,
  K.~Nishiyama, M.~Horvatic, and C.~Bertheir: Prog. Theor. Phys. Suppl.
  {\bfseries 159} (2005) 1.

\bibitem{JOK:PRL11}
H.~Jeschke, I.~Opahle, H.~Kandpal, R.~Valenti, H.~Das, T.~Saha-Dasgupta,
  O.~Janson, H.~Rosner, A.~Br{\"u}hl, B.~Wolf, M.~Lang, J.~Richter, S.~Hu,
  X.~Wang, R.~Peters, T.~Pruschke, and A.~Honecker: Phys. Rev. Lett. {\bfseries
  106} (2011) 217201.

\bibitem{ONO:NC19}
R.~Okuma, D.~Nakamura, T.~Okubo, A.~Miyake, A.~Matsuo, K.~Kindo, M.~Tokunaga,
  N.~Kawashima, S.~Takeyama, and Z.~Hiroi: Nat. Commun. {\bfseries 10} (2019)
  1229.

\bibitem{FMM:NC20}
M.~Fujihala, K.~Morita, R.~Mole, S.~Mitsuda, T.~Tohyama, S.-i. Yano, D.~Yu,
  S.~Sota, T.~Kuwai, A.~Koda, H.~Okabe, H.~Lee, S.~Itoh, T.~Hawai, T.~Masuda,
  H.~Sagayama, A.~Matsuo, K.~Kindo, S.~Ohira-Kawamura, and K.~Nakajima: Nat.
  Commun. {\bfseries 11} (2020) 3429.

\bibitem{Har:21}
M.~Hardtke: Bachelor thesis, Bielefeld University, Faculty of Physics (2021).

\end{thebibliography}

\appendix

\section{Constructing a density of states}
\label{sec-A}
Through a suitable shift of the energy scale ($E_0 > 0)$  the partition function can be written as the Laplace transform of the density of states:
\begin{align}
    Z(\beta) = \int_0^\infty \rho(E)e^{-\beta E}\text{d}E \quad 
\end{align}
with 
\begin{align}
\rho(E) = \sum_n \delta(E-E_n)
\end{align}
Hence an inverse Laplace transformation of the partition function approximated by FTLM will give an approximation of the density of states:
\begin{align}
    \rho_{\text{FTLM}} = \frac{1}{2\pi \textbf{i}}\int_{c-\i\infty}^{c+\textbf{i}\infty}Z_{\text{FTLM}}(\beta)e^{\beta E}\text{d}\beta\ .
\end{align}
By substituting $\beta$ with $c + \i s$ one finds:
\begin{align}
    &\rho_{\text{FTLM}}(E)    = 
                                \frac{\textbf{i}}{2\pi \textbf{i}}
                                \int_{-\infty}^{\infty}
                                Z\subt{FTLM}(c + \textbf{i} s)
                                e^{(c+\textbf{i} s) E}
                                \text{d}s\\
                            & = 
                                \frac{1}{2 \pi}
                                \sum_{\Gamma, n, r}
                                \gamma_n^{(r, \Gamma)}
                                \int_{-\infty}^{\infty}
                                e^{-(c+ \i s) \epsilon^{(r,\Gamma)}_n}
                                e^{(c+\i s) E}
                                \text{d}s\\
                            & = 
                                \sum_{\Gamma, n, r}
                                \gamma_n^{(r, \Gamma)}
                                e^{-c(\epsilon^{(r,\Gamma)}_n-E)}
                                \frac{1}{2 \pi}
                                \int_{-\infty}^{\infty}
                                e^{-\i s (\epsilon^{(r,\Gamma)}_n-E)}
                                \text{d}s\\
                            & =
                                \sum_{\Gamma, n, r}
                                \gamma_n^{(r, \Gamma)}
                                \delta(\epsilon^{(r,\Gamma)}_n-E)\ .
\end{align}
The difficulty is now to find a suitable representation of the $\delta$-distribution, which will give a good representation of the density of states as well. For (energetically) bounded systems a rectangular function
\begin{align}
    \delta_{\varepsilon}(x) =   \begin{cases}
                                \frac{1}{\varepsilon}  & \ \text{if}\  |x|\leq\frac{\varepsilon}{2}\\
                                0                   & \ \text{else}
                                \end{cases}
\end{align}
is convenient because it is also a bounded function and therefore will not give non-zero values outside the true spectrum.
An additional problem of the representation is that the number of pseudo-eigenvalues obtained 
from the FTLM approximation
is much smaller than the dimension of the Hilbert space. 
So if $\varepsilon$ is chosen too small it is possible to produce gaps where in the exact density of states there are no gaps. 
This issue is of course only relevant in very dense parts of the density.  
As a solution one can choose the bounds of the rectangular functions to always 
be the mean of to consecutive pseudo-eigenvalues so that all gaps will be closed
\cite{Har:21}.
For this one defines an asymmetric version of the rectangular function:
\begin{align}
    \delta_{\varepsilon,\,\varepsilon'}(x) =   
        \begin{cases}
            \frac{1}{\varepsilon + \varepsilon'}   & \ \text{if}\  -\varepsilon' \leq x\leq \varepsilon\\
            0                       & \ \text{else}
        \end{cases}\ .
\end{align}
 the parameters will chosen such that the bounds will lie at mean of to consecutive but non-degenerate pseudo-eigenvalues. 
 Let $\epsilon_k^\Gamma$ be the sorted but pairwise distinct pseudo-eigenvalues of a subspace $\Gamma$, then the $\delta$-distributions can be replaced by
\begin{align}
    \delta(\epsilon^{\Gamma}_k-E) \rightarrow \delta_{\varepsilon_{+},\,\varepsilon_{-}}(\epsilon^{\Gamma}_k-E) \quad 
\end{align}
    with 
\begin{align}
        \varepsilon_{-}= \tfrac{\epsilon^{\Gamma}_{k}-\epsilon^{\Gamma}_{k-1}}{2},\ 
        \varepsilon_{+} = \tfrac{\epsilon^{\Gamma}_{k+1}-\epsilon^{\Gamma}_{k}}{2}\ .
\end{align}
Lanczos weights of (numerically) degenerate pseudo-eigenvalues will be binned and added, 
where $N\subt{dis}$ denotes the number of pseudo-eigenvalues without the dropped duplicates.
This gives rise to the following representation of the density of states:
\begin{align}
    \rho_{\text{FTLM}}(E) \approx
        \sum_{\Gamma}
        \sum^{N\subt{dis}}_{k=1}
        \gamma_k^{\Gamma}
        \delta_{\varepsilon_{+},\,\varepsilon_{-}}(\epsilon^{\Gamma}_k-E)\ .
\end{align}

\end{document}